\documentclass[showpacs,eqsecnum,superscriptaddress,nofootinbib,preprint,11pt]{revtex4}

\usepackage{amsmath} 
\usepackage{amssymb} 

\begin{document}

\title{Quantum Gravity: A Brief History of Ideas and Some Prospects\footnote{This article is to be published in International Journal of Modern Physics D and also in the book ``One Hundred Years of General Relativity: From Genesis and Empirical Foundations to Gravitational Waves, Cosmology and Quantum Gravity,'' edited by Wei-Tou Ni (World Scientific, Singapore, 2015).}}

\author{Steven Carlip}
\email{carlip@physics.ucdavis.edu}
\affiliation{Physics Department, University of California at Davis, Davis, CA 95616, USA}

\author{Dah-Wei Chiou}
\email{dwchiou@gmail.com}
\affiliation{Department of Physics, National Taiwan Normal University, Taipei 11677, Taiwan}
\affiliation{Department of Physics and Center for Condensed Matter Sciences, National Taiwan University, Taipei 10617, Taiwan}

\author{Wei-Tou Ni}
\email{weitou@gmail.com}
\affiliation{Center for Gravitation and Cosmology, Department of Physics,
National Tsing Hua University, Hsinchu 30013, Taiwan}

\author{Richard Woodard}
\email{woodard@phys.ufl.edu}
\affiliation{Department of Physics, University of Florida, Gainesville, FL 32611, USA}

\begin{abstract}
We present a bird's-eye survey on the development of fundamental ideas of quantum gravity, placing emphasis on perturbative approaches, string theory, loop quantum gravity, and black hole thermodynamics. The early ideas at the dawn of quantum gravity as well as the possible observations of quantum gravitational effects in the foreseeable future are also briefly discussed.
\end{abstract}



\maketitle

\section{Prelude}

Quantum gravity is the research that seeks a consistent unification of the two foundational pillars of modern physics --- quantum theory and Einstein's theory of general relativity. It is commonly considered as the paramount open problem of theoretical physics, and many fundamental issues --- such as the microscopic structure of space and time, the origin of the universe, the resolution to spacetime singularities, etc. --- relies on a better understanding of quantum gravity.

The quest for a satisfactory quantum description of gravity began very early. Einstein after proposing general relativity thought that quantum effects must modify general relativity in his first paper on gravitational waves in 1916 \cite{Einstein:1916} (although he switched to a different point of view working on the unification of electromagnetism and gravitation in the 1930s). Klein argued that the quantum theory must ultimately modify the role of spatio-temporal concepts in fundamental physics \cite{Klein:1927,Klein:1954,Klein:1956} and his ideas were developed by Deser \cite{Deser:1957zz}. With the interests and developments of Rosenfeld, Pauli, Blokhintsev, Heisenberg, Gal'perin, Bronstein, Frenkel, van Dantzig, Solomon, Fierz, and many other researchers, three approaches to quantum gravity after World War II had already been enunciated in 1930s pre-World War II as summarized by Stachel \cite{Stachel:1916}:
\begin{enumerate}
\item Quantum gravity should be formulated by analogy with quantum electrodynamics (Rosenfeld, Pauli, Fierz).
\item The unique features of gravitation will require special treatment --- the full theory with its nonlinear field equations must be quantized and generalized as to be applicable in the absence of a background metric (Bronstein, Solomon).
\item General relativity is essentially a macroscopic theory, e.g.\ a sort of thermodynamics limit of a deeper, underlying theory of interactions between particles (Frenkel, van Dantzig).
\end{enumerate}
Many ideas of the early time continue to provide valuable insight about the nature of quantum gravity even today. For example, the 1939 work on linearized general relativity as spin-2 field by Fierz and Pauli \cite{Fierz:1939ix} inspired a recent development on massive gravity, bimetric gravity, etc.

Modern work on quantum gravity, however, did not really start off until the development of a canonical formalism in 1959-1961 by Arnowitt, Deser, and Misner (ADM) for the case of asymptotically flat boundary conditions \cite{Deser:1959zza, Arnowitt:1959ah, Arnowitt:1960es, Deser:1960zza, Arnowitt:1960zzc, Deser:1960zzc, Arnowitt:1960zza, Deser:1960zzb, Deser:1961zza, Arnowitt:1961zza, Arnowitt:1961zz} (for a review, see Ref.~\cite{Arnowitt:1962hi}). This served as the basis for Schoen and Yau's proof that the classical energy is bounded below \cite{Schoen:1979rg,Schoen:1981vd} and is the starting point for numerical simulations of classical general relativity. Of course the ADM formalism also provides the Hamiltonian and canonical variables whose quantization defines quantum general relativity. However, the sacrifice of manifest covariance made perturbative computations prohibitively difficult, although many strategies to tackle this difficulty have been suggested e.g.\ in loop quantum gravity.

Ninety-nine years passed since the very first conception of Einstein. Today, quantum gravity has grown into a vast area of research in many different (both perturbative and nonperturbative) approaches.
Many directions have led to significant advances with various appealing and ingenious ideas; these include causal sets, dynamical triangulation, emergent gravity, $\mathcal{H}$ space theory, loop quantum gravity, noncommutative geometry, string theory, supergravity, thermogravity, twistor theory, and much more. (For surveys and references on various approaches of quantum gravity, see Refs.~\cite{Rovelli:1997qj,Carlip:2001wq,Woodard:2009ns,Kiefer:2012boa}; for detailed accounts of the history and development, see Refs.~\cite{Stachel:1916,Rovelli:2000aw}; for a popular science account of quantum gravity, see the book Ref.~\cite{Smolin:2000af}.)


\section{Perturbative quantum gravity}

The quantization of general relativity on the perturbative level
(with or without matter) was the life work of DeWitt, for
which he was awarded the 1987 Dirac Medal and the 2005 Einstein
Prize \cite{DeWitt:1967yk,DeWitt:1967ub,DeWitt:1967uc}. His program
eventually succeeded --- against the expectation of most expert
opinion during the 1960s --- in generalizing Feynman's covariant
quantization of quantum electrodynamics so that it could be applied
as well to theories such as general relativity and Yang-Mills that
are based on non-Abelian gauge symmetries. There were three key
steps:
\begin{enumerate}
\item{The introduction of the {\it background field method} for
representing the effective action as a gauge invariant functional
of the fields \cite{DeWitt:1967ub,Abbott:1981ke}. This made clear
the connection between invariant counterterms and all possible
ultraviolet divergences of scattering amplitudes at a fixed order
in the loop expansion.}
\item{The realization that non-Abelian gauge symmetries require
the inclusion of opposite-normed {\it ghost fields} to compensate
the effect of unphysical polarizations in loop corrections
\cite{DeWitt:1967ub,Faddeev:1967fc}.}
\item{The development of {\it invariant regularization techniques},
the first of which was dimensional regularization
\cite{'tHooft:1972fi,Bollini:1972ui}. This permitted efficient
computations to be made without the need for non-invariant
counterterms.}
\end{enumerate}

The first application of these techniques was made in 1974 by 't
Hooft and Veltman \cite{'tHooft:1974bx}. They showed that general
relativity without matter has a finite $S$-matrix at one loop order,
but that the addition of a scalar field leads to the need for
higher curvature counterterms that would destabilize the universe.
Very shortly thereafter Deser and van Nieuwenhuizen showed that
similar unacceptable counterterms are required at one loop for
general relativity plus Maxwell's electrodynamics
\cite{Deser:1974zzd,Deser:1974cz}, for general relativity plus a
Dirac fermion \cite{Deser:1974cy}, and (with Tsao) for general
relativity plus Yang-Mills \cite{Deser:1974xq}. The fate of pure
general relativity was settled in 1985 when Goroff and Sagnotti
showed that higher curvature counterterms are required at two
loop order \cite{Goroff:1985sz,Goroff:1985th,vandeVen:1991gw}.

These results show that quantum general relativity is not
fundamentally consistent as a perturbative quantum field theory.
Opinion is divided as to whether this means that general relativity
must be abandoned as the fundamental theory of gravity or whether
quantum general relativity might make sense nonperturbatively.
String theory discards general relativity, and does not even
employ the metric as a fundamental dynamical variable. Loop quantum gravity and causal dynamical triangulation are approaches that
attempt to make sense of quantum general relativity without employing
perturbation theory. At this time it would be fair to say that there
is not yet any fully successful quantum theory of gravity.

But being unable to do {\it everything} is not at all the same as
lacking the ability to do {\it anything}. Perturbative quantum general
relativity (with or without matter) can still be used in the standard
sense of low energy effective field theory \cite{Donoghue:1993eb,
Donoghue:1994dn}, in the same way that Fermi theory was for years
employed to understand the weak interactions, even including loop
effects \cite{Feinberg:1968zz,Hsu:1992tg}. The most interesting
effects tend to occur in curved spacetime backgrounds, on which one
can also consider quantizing only matter, without dynamical gravity
\cite{DeWitt:1975ys}.

When working on curved backgrounds one often finds the formalism of
flat space scattering theory to be inappropriate because it is based
on incorrect assumptions about free vacuum at very early times (which
actually began with an initial singularity) and at very late times
(which may be filled with ensembles of particles created by the
curved geometry). Solving the effective field equations under these
conditions is especially troublesome because the in-out effective
field equations at a spacetime point $x^{\mu}$ contain contributions
from points ${x'}^{\mu}$ far in its future, and because even the
in-out matrix elements of Hermitian operators such as the metric and
the Maxwell field strength tensor can develop imaginary parts. For
these reasons it is often more appropriate to employ the
Schwinger-Keldysh formalism. This is a diagrammatic technique that
is almost as simple to use as the standard Feynman rules, which gives
true expectation values instead of in-out matrix elements. It was
devised for quantum mechanics in 1960 by Julian Schwinger
\cite{Schwinger:1960qe}. Over the next few years it was generalized
to quantum field theory by Mahanthappa and Bakshi
\cite{Mahanthappa:1962ex,Bakshi:1962dv,Bakshi:1963bn}, and to
statistical field theory by Keldysh \cite{Keldysh:1964ud}. Although
Schwinger-Keldysh effective field equations are nonlocal, they are
causal --- in the sense that only points ${x'}^{\mu}$ in the past of
$x^{\mu}$ contribute --- and the solutions for Hermitian fields are
real \cite{Chou:1984es,Jordan:1986ug}. They are the natural way to
study quantum effects in cosmology \cite{Calzetta:1986ey} and
nonequilibrium quantum field theory \cite{Calzetta:1986cq}. Steven
Weinberg recently devised a variant of the formalism which is
especially adapted to computing the correlation functions of
primordial inflation \cite{Weinberg:2005vy,Weinberg:2006ac}.

For a recent review of perturbative quantum gravity, see Ref.~\cite{Woodard:2014jba}

\section{String theory}

On the other hand, the search for a completely consistent theory of quantum gravity has never stopped. Although it does not embrace general relativity at the foundational level and it is disputable whether it can be genuinely formulated in a background-independent, nonperturbative fashion, string theory is arguably considered by many as the most promising candidate for a consistent theory of quantum gravity on the grounds that its low-energy limit surprisingly gives rise to (the supersymmetric extension of) the theory of (modified) gravity plus other force and matter fields. In the framework of string theory, the point-like particles of particle physics are replaced by (the different quantum excitations of) minuscule one-dimensional objects called \emph{strings}. In addition to being a quantum theory of gravity, string theory also aims to unify all fundamental forces and all forms of matter, striving for the ultimate goal of being a ``theory of everything''. (For textbooks on string theory, see Refs.~\cite{Green:1987sp,Polchinski:1998rq,Becker:2007zj}; for a detailed account of its history and development, see Ref.~\cite{Rickles:2014fha}.)

String theory was born in the late 1960s as a never completely successful theory of the strong nuclear force and was later recognized as a suitable framework for a quantum theory of gravity (see \cite{Schwarz:2007yc} for its ``prehistory''). The idea of identifying the string action as the area of the worldsheet of the string traveling in spacetime was introduced independently by Nambu \cite{Nambu:1986ze}, Goto \cite{Goto:1971ce}, and Hara \cite{Hara:1971ur} in the early 1970s. The modern treatment of string theory based on the worldsheet path integral of the Polyakov action was initiated by Polyakov in 1981 \cite{Polyakov:1981rd,Polyakov:1981re} and has led to an intimate link with the conformal field theory. The worldsheet conformal field theory in the presence of spacetime background fields was studied by Callan \textit{et al.}\ \cite{Callan:1985ia}; the conformal invariance demands beta functions of the worldsheet field theory to vanish identically and consequently yields the equations of motion of the background fields that bear remarkable resemblance to the Einstein field equation and (non-abelian) Maxwell's equations with higher-order corrections.

Quantization of the bosonic string requires the number of spacetime dimensions to be 26, of which 22 extra spatial dimensions are thought to be compactified in the deep microscopic scale and thus undetectable at low energies. The bosonic string theory is unsatisfactory in two aspects: first, it entails existence of negative-normed tachyon fields and consequently is non-unitary and inconsistent; second, it does not contains fermions and thus cannot account for the quarks and leptons in the standard model. In the 1980s, bosonic string theory was extended into superstring theory, which incorporates both bosonic and fermionic degrees of freedom via supersymmetry and requires 6 extra spatial dimensions to be compactified on a Calabi-Yau manifold. Superstring theory not only gets rid of negative-normed fields but also includes fermions as desired. There are basically two (equivalent) approaches to embody supersymmetry in string theory: (i) the Ramond-Neveu-Schwarz (RNS) formalism with manifest supersymmetry on the string worldsheet \cite{Ramond:1971gb,Neveu:1971rx}; (ii) the Green-Schwarz (GS) formalism with manifest supersymmetry on the background spacetime \cite{Green:1980zg,Green:1981xx,Green:1981yb}. The \emph{first superstring revolution} began in 1984 with the discovery by Green and Schwarz \cite{Green:1984sg} that the cancellation of gauge and gravitational anomalies demands a very strong constraint on the gauge symmetry: the gauge group must be either $SO(32)$ or $E_8\times E_8$. Eventually, five consistent but distinct superstring theories were found: type I, type IIA, type IIB, $SO(32)$ heterotic, and $E_8\times E_8$ heterotic.

In the late 1980s, it was realized that the two type II theories and the two heterotic theories are related by T-duality, which, simply speaking, maps a string winding around a compactified dimension of radius $R$ to that of radius $\ell_\mathrm{s}^2/R$, where $\ell_\mathrm{s}$ is the string length scale. (For reviews on T-duality, see Refs.~\cite{Giveon:1994fu,Alvarez:1994dn}.) A few years later, S-duality was discovered as another kind of duality that maps the string coupling constant $g_\mathrm{s}$ to $1/g_\mathrm{s}$. The two basic examples are the duality that maps the type I theory to the $SO(32)$ heterotic theory \cite{Polchinski:1995df} and the duality that maps the type IIB theory to itself \cite{Hull:1994ys}. As S-duality relates a strongly coupled theory to a weakly coupled theory, it provides a powerful tool to explore nonperturbative behaviors of a theory with $g_\mathrm{s}>1$, given the knowledge of the dual theory perturbatively obtained with $g_\mathrm{s}<1$. In the mid-1990s, understanding of the nonperturbative physics of superstring theory progressed significantly, revealing that superstring theory contains various dynamical objects with $p$ spatial dimensions called $p$-branes, in addition to the fundamental strings. Particularly, type I and II theories contain a class of $p$-branes, known as D-branes (or D$p$-branes, more specifically), upon which open strings can end in Dirichlet boundary conditions. In 1995, Polchinski identified D-branes as solitonic solutions of supergravity that are understood as (generalized) black holes \cite{Polchinski:1995mt}, a landmark discovery that heralded the prominence of D-brane dynamics.

Furthermore, studies on the S-duality of type II \cite{Townsend:1995kk,Witten:1995ex} and $E_8\times E_8$ heterotic \cite{Horava:1995qa,Horava:1996ma} theories sprang a big surprise: these two theories grow an eleventh dimension of size $g_\mathrm{s}\ell_\mathrm{s}$ at strong coupling. These discoveries together with the aforementioned relations between different superstring theories via T- and S-dualities brought about the \emph{second superstring revolution} that took place around 1994--1997, suggesting the existence of a more fundamental theory in 11 dimensions. The existence of such a theory was first conjectured and named as \emph{M-theory} by Witten at a string theory conference at University of Southern California in 1995. M-theory is supposed to be \emph{the} fundamental theory of everything, of which the five distinct superstring theories along with the 11-dimensional theory of supergravity are regarded as different limits. Although the complete formulation of M-theory remains elusive (and whether ``M'' should stand for ``mother'', ``magic'', ``mystery'', ``membrane'', ``matrix'', etc.\ remains a matter of taste), a great deal about it can be learned by virtue of the fact that the theory should describe 2- and 5-dimensional objects known as M-branes (M2- and M5-branes, respectively) and reduce to 11-dimensional supergravity theory at the low-energy limit. Investigations of M-theory have inspired a great number of important theoretical results in both physics and mathematics.

A collection of coincident D-branes in string theory or M-branes in M-theory produces a warped spacetime with flux of gauge fields akin to a charged black hole, as the branes are sources of gauge flux and gravitational curvature. The low-energy limit of the gauge theory on the branes' worldvolume (referred to as ``boundary'') is found to describe the same physics of string theory or M-theory in the near-horizon geometry (referred to as ``bulk''). In this way, one is led to conjecture a remarkable duality that relates conventional (non-gravitational) quantum field theory on the boundary to string theories or M-theory on the bulk. This gauge theory/string theory duality is often referred to as anti-de Sitter/conformal field theory (AdS/CFT) correspondence, which was first spelled out in a seminal paper by Maldacena in 1997 \cite{Maldacena:1997re}. (A detailed review on AdS/CFT correspondence was given in Ref.~\cite{Aharony:1999ti}.) By considering $N$ coincident D3-branes in the type IIB theory, one obtains the celebrated example: the type IIB theory in $AdS_5\times S^5$ (the product space of 5-dimensional anti-de Sitter space and 5-dimensional sphere) is equivalent to the $SU(N)$ super Yang-Mills theory with $\mathcal{N}=4$ supersymmetry on the 4-dimensional boundary (which is a super-conformal field theory). Many other examples have also been substantiated to various degrees of rigor. In the attempt to construct the complete formulation of M-theory, the AdS/CFT correspondence sets out a new strategy based on the \emph{holographic principle}, which posits that the physics of a bulk region is completely encoded on its lower-dimensional boundary, as originally propounded by 't~Hooft in 1993 \cite{'tHooft:1993gx} and elaborated by Susskind in 1994 \cite{Susskind:1994vu}.

In string theory, the shape and size of the compactified manifold and consequently the fundamental physical constants (lepton masses, coupling constants, cosmological constant, etc.) are dynamically determined by vacuum expectation values of scalar (moduli) fields. The crucial question arises known as the \emph{moduli-space problem} or \emph{moduli-stabilization problem}: what mechanism stabilizes the compactified manifold and uniquely determines the physical constants? Answers to this problem have been proposed in the approach of \emph{flux compactifications} as a generalization of conventional Calabi-Yau compactifications, which where first introduced by Strominger in 1986 \cite{Strominger:1986uh} and by de~Wit, Smit, and Hari Dass\ in 1987 \cite{de Wit:1986xg} and now have become a rapidly developing area of research. The idea is to generate a potential that stabilizes the moduli fields by compactifying string theory or M-theory on a \emph{warped geometry}. In a warped geometry, the fluxes associated with certain tensor fields thread cycles of the compactified manifold; as the magnetic flux of an $n$-form field strength through an $n$-cycle depends only on the homology of the $n$-cycle, the flux stabilizes due to the flux quantization condition and gives rise to a nonvanishing potential. Flux compactifications were first studied in the context of M-theory by Becker and Becker in 1996 \cite{Becker:1996gj}. (A comprehensive review on flux compactifications was given in Ref.~\cite{Grana:2005jc}.)

In 1999, Gukov, Vafa, and Witten made it evident that flux compactfications generate nonvanishing potential for moduli fields, leading to a solution to the moduli-space problem \cite{Gukov:1999ya}. However, since the fluxes can take many different discrete values over different homology cycles of different (generalized) Calabi-Yau manifolds, flux compactifications typically yield a vast multitude (commonly estimated to be of the order $10^{500}$  \cite{Douglas:2003um}) of possible vacuum expectation values, referred to as the \emph{string theory landscape}, and therefore we have to abandon the long-cherished hope that the fundamental physical constants are supposed to be uniquely fixed by string theory or M-theory. In response to this problem, Susskind in 2003 \cite{Susskind:2003kw} proposed the anthropic argument of the string theory landscape as a concrete implementation of the ``anthropic principle'', which suggests that physical constants take their values not because fundamental laws of physics dictate so but rather because such values are at the loci of the landscape that are suitable to the existence of (intelligent) life (in order to measure these constants). The properties of the string theory landscape was shortly analyzed in a statistical approach by Douglas \cite{Douglas:2003um}.
The scientific relevance of the anthropic landscape has sparked fierce debate and remained highly controversial, yet it has gradually gained popularity (especially in the context of cosmology).

The application of flux compactifications as well as the string theory landscape to the study of cosmology has spawned a whole new growing research area known as \emph{string cosmology}. A notable subfield of it is \emph{brane cosmology}, which posits that our four-dimensional universe is restricted to a brane inside a higher-dimensional space, as first proposed by Randall and Sundrum in 1999 \cite{Randall:1999vf}. The primary goal of string cosmology is to place field-theoretic models of cosmological inflation on a firmer logical ground from the perspective of string theory or, more boldly, to open entirely new realms of pre-big bang scenarios that cannot be described by any field-theoretic models. In 2003, Kachru \textit{et al.}\ outlined the construction of metastable de Sitter vacua \cite{Kachru:2003aw} and shortly investigated the application to inflation \cite{Kachru:2003sx}. String cosmology has made some concrete predictions that will soon be confronted with the near-future astrophysical observations. (For reviews on string cosmology, see Refs.~\cite{Quevedo:2002xw,Danielsson:2004mf} and the textbook Ref.~\cite{Baumann:2014nda}.)

\section{Loop quantum gravity}\label{sec:LQG}

The primary competitor of string theory for a complete quantum theory is loop quantum gravity (LQG). In sharp contrast to the ambitious goal of string theory, LQG does not intend to pursue a theory of everything that unifies all force and matter fields but deliberately adopts the ``minimalist'' approach in the sense that it focuses solely on the search for a consistent quantum theory of gravity without entailing any extraordinary ingredients such as extra dimensions, supersymmetry, etc.\ (although many of these can be incorporated compatibly). The beauty of LQG lies in its faithful attempt to establish a conceptual framework whereby the apparently conflicting tenets of quantum theory and Einstein's theory of general relativity conjoin harmoniously, essentially in a nonperturbative, background-independent fashion.
(For textbooks on LQG, see Refs.~\cite{Rovelli:2004tv,Thiemann:2007zz,Rovelli:2014book}; for a detailed account of its history and development, see Ref.~\cite{Rovelli:2010bf}; for a recent review of LQG, see Ref.~\cite{Chiou:2014jwa}.)

The research of LQG originated from the works by Ashtekar in 1986--1987 \cite{Ashtekar:1986yd,Ashtekar:1987gu}, which reformulated Einstein's general relativity into a new canonical formalism in terms of a selfdual spinorial connection and its conjugate momentum, now known as \emph{Ashtekar variables}, casting general relativity in a language closer to that of the Yang-Mills gauge theory. The shift from metric to Ashtekar variables provided the possibility of employing nonperturbative techniques in gauge theories and opened a new avenue eventually leading to LQG.
Shortly thereafter, Jacobson and Smolin discovered that Wilson loops of the Ashtekar connection are solutions of the Wheeler-DeWitt equation (the formal equation of quantum gravity) \cite{Jacobson:1987qk}. This remarkable discovery led to the ``loop representation of quantum general relativity'' introduced by Rovelli and Smolin in 1988-1990 \cite{Rovelli:1987df,Rovelli:1989za}. The early developments of LQG were reviewed in Ref.~\cite{Rovelli:1991zi}.

In the 1990s, by the works of Br\"{u}gmann, Gambini, and Pullin \cite{Bruegmann:1992ak,Bruegmann:1992gp}, it became clear that intersections of loops are essential for the consistency of the theory and quantum states of gravity should be formulated in terms of intersecting loops, i.e., \emph{graphs} with links and nodes. Later on, inspired by Penrose's speculation on the combinatorial structure of space \cite{Penrose:1971}, Rovelli and Smolin \cite{Rovelli:1995ac} obtained an explicit basis of states of quantum geometry known as \emph{spin networks}, which are oriented graphs labelled by numbers associated with spin representations of $SU(2)$ on each link and node. The mathematical construction of spin networks was systematized by Baez \cite{Baez:1994hx,Baez:1995md}. The idea that LQG could predict discrete quantum geometry was first suggested in the study of ``weave states'' \cite{Ashtekar:1992tm}. In 1995, the area and volume operators were defined upon spin networks and their eigenvalue spectra were found to be discrete \cite{Rovelli:1994ge}, showing that quantum geometry is indeed quantized in the Planck scale as had long been speculated. Rigourous and systematic studies of geometry operators and their eigenvalues were further elaborated (e.g. see Refs.~\cite{De Pietri:1996pja,Thiemann:1996au,Lewandowski:1996gk}).

The physical interpretation of spin networks (or more precisely, diffeomorphism-invariant knot classes of spin networks, known as \emph{s-knots}) is extremely appealing: they represent different quantized 3-dimensional geometries, which are not quantum excitations \emph{in} space but \emph{of} space. This picture manifests the paradigm of background independence of general relativity, as any reference to the localization of spin networks is dismissed. Furthermore, inclusion of matter fields into LQG does not require a major revamp of the underlying framework. (For a systematic account of inclusion of matter, see Ref.~\cite{Thiemann:2007zz}.) The quantum states of space plus matter naturally extend the notion of spin networks with additional degrees of freedom. In the presence of matter fields, the background independence becomes even more prominent, as geometry and matter fields are genuinely on the equal footing and reside on top of one another via their contiguous relations in the spin network without any reference to a given background.

While kinematics of LQG is well understood in terms of spin networks, its dynamics (i.e., evolution of spin networks) and low-energy (semi-classical) physics is much more difficult and unclear. An anomaly-free formulation of the quantum dynamics was first obtained by Thiemann in 1996 \cite{Thiemann:1996ay}, and shortly the formulation was fully developed in his remarkable ``QSD'' series of papers \cite{Thiemann:1996aw,Thiemann:1996av,Thiemann:1997rv}. Since then, a great number of variant approaches have been pursued, but the quantum dynamics remains very obscure, largely because implementation of the quantum Hamiltonian constraint is very intricate. The \emph{Master constraint program} \cite{Thiemann:2003zv} initiated by Thiemannn proposed an elegant solution to the difficulties and has evolved into a fully combinatorial theory known as \emph{algebraic quantum gravity} (AQG) \cite{Giesel:2006uj,Giesel:2006uk,Giesel:2006um,Giesel:2007wn}. Meanwhile, various techniques, notably by the idea of \emph{holomorphic
coherent states} \cite{Ashtekar:1994nx, Thiemann:2002vj, Bahr:2007xa, Bahr:2007xn, Flori:2009rw, Bianchi:2009ky}, have been devised to investigate the low-energy physics. Up to now, LQG has achieved considerable progress in understanding the quantum dynamics and providing contact with the low-energy physics.

Applying principles of LQG to cosmological settings leads to the symmetry-reduced theory known as loop quantum cosmology (LQC), which was originally proposed by Bojowald in 1999 and reached a rigourous formulation around 2006. Providing a ``bottom-up'' approach to the full theory of LQG, LQC has become the most well-developed subfield of LQG and led to many significant successes. Most notably, it suggests a new cosmological scenario where the big bang is replaced by the \emph{quantum bounce}, which bridges the present expanding universe with a preexistent contracting counterpart. Absence of cosmological singularities has been shown to be robust for a great variety of LQC models, therefore affirming the long-held conviction that singularities in classical general relativity should be resolved by the effects of quantum gravity. (For textbooks and reviews on LQC, see Refs.~\cite{Bojowald:2008zzb, Ashtekar:2011ni, Bojowald:2011zzb, Bojowald:2011book}.)

The standard formalism of LQG adopts the canonical (Hamiltonian) approach but is closely related to the covariant (sun-over-histories) approach of quantum gravity known as the \emph{spin foam} theory. The concept of a spin foam, which can be viewed as a (discrete) ``worldsurface'' swept out by a spin netwrok traveling and transmuting in time, was introduced in 1993 as inspired by the work of Ponzano-Regge model \cite{LaFave:1993zp} and later developed into a systematic framework of quantum gravity by Perez \textit{et al}.\ in the 2000s. A spin foam represents a quantized spacetime in the same sense that a spin network represents a quantized space; the transition amplitude from one spin network to another is given as the discrete sum (with appropriate weights) over all possible spin foams that connect the initial and final spin networks.
(For reviews on the spin foam theory, see Refs.~\cite{Perez:2003vx, Oriti:2003wf, Perez:2004hj, Perez:2012wv}.)

LQG has grown into a very active research field pursued in many directions, both in the traditional canonical approach and in the covariant (i.e.\ spin foam) approach. Recently, the precise connection between the canonical and covariant approaches has become one of the central topics of LQG. The merger of different approaches has yielded profound insight and suggested a new theoretical framework referred to as \emph{covariant loop quantum gravity}, which provides new conceptual principles and could pave a royal road to a complete theory of quantum gravity as advocated by Rovelli \cite{Rovelli:2010wq,Rovelli:2011eq,Rovelli:2013ht} (also see Ref.~\cite{Rovelli:2014book} for a comprehensive account).

\section{Black hole thermodynamics}
Perhaps the most notable achievement in the study of quantum gravity so far was the discovery that black holes are not really black, but emit thermal \emph{Hawking radiation}. The surprising fact that black holes behave as thermodynamic objects has radically affected our understanding of general relativity and given valuable hints about the nature of quantum gravity.

The study of black hole thermodynamics can be traced back to 1972 when Hawking proved that an area of an event horizon can never decrease \cite{Hawking:1971vc} --- a property reminiscent of the second law of thermodynamics. The resemblance between black hole mechanics and thermodynamics was further enhanced by the discovery of analogs of other thermodynamical laws and formally elaborated in the paper by Bardeen, Carter, and Hawking \cite{Bardeen:1973gs} in terms the ``four laws of black hole mechanics'', which are in clear parallel with the four usual laws of thermodynamics as the surface gravity ($\kappa$) plays the role of temperature and the area of horizon ($A_\mathrm{hor}$) of entropy. Around the same time, considering a series of thought experiments initiated by his advisor, Wheeler \cite{Wheeler's cup of tea}, Bekenstein argued that the black hole entropy has to be of the form $S_\mathrm{BH}=\eta k_\mathrm{B}A_\mathrm{hor}/(\hbar G)$, where $k_\mathrm{B}$ is Boltzmann's constant and $\eta$ is a proportional constant of order one, and correspondingly obtained the ``generalized second law of thermodynamics'', which asserts that the total of ordinary entropy of matter plus black hole entropy never decreases \cite{Bekenstein:1972tm,Bekenstein:1973ur,Bekenstein:1974ax}.

In 1974, using the newly developed techniques of quantum field theory in curved spacetime, Hawking \cite{Hawking:1974rv,Hawking:1974sw} showed that \emph{all} black holes radiate as black bodies with the Hawking temperature $T_\mathrm{H}=\hbar\kappa/(2\pi)$. The first law of black hole mechanics then determines the entropy as $S_\mathrm{BH}=\frac{1}{4}k_\mathrm{B}A_\mathrm{hor}/(\hbar G)$, which confirms Bekenstein's expression with $\eta=1/4$ and is often referred to as the Bekenstein-Hawking formula. As the equivalence principle implies that the gravitational field near a black hole horizon is locally equivalent to uniform acceleration in a flat spacetime, it is expected that an accelerated observer should perceive an effect similar to the Hawking radiation. In 1976, Unruh \cite{Unruh:1976db} demonstrated that an observer moving with a constant proper acceleration $a$ in Minkowski spacetime indeed see a thermal flux of particles with the Unruh temperature $T_\mathrm{U}=\hbar a/(2\pi)$, which is in almost exact analogy with the Hawking temperature. At almost the same time, Bisognano and Wichmann gave an independent and mathematically rigorous proof of the Unruh effect in quantum field theory \cite{Bisognano:1976za}. To resolve the worrying issue whether the flux of particles \emph{formally} obtained via the standard quantum field-theoretical definitions of vacuum states and particle numbers can be \emph{physically} interpreted as ``particles'' as seen by detectors, Unruh \cite{Unruh:1976db} and DeWitt \cite{DeWitt:Quantum Gravity} devised simple models of particle detectors and showed that the answer is affirmative for these detectors.

The Hawking temperature and the Bekenstein-Hawking entropy are inherently quantum gravitational, in the sense that they depend explicitly on both Planck's constant $\hbar$ and Newton's constant $G$.  They are also surprisingly universal.  The entropy, for example, depends only on the horizon area, and takes the same simple form regardless of the black hole's charges, angular momentum, horizon topology, and even the number of spacetime dimensions.
The thermal properties of black holes have not been directly observed, but by now they have been derived in so many different ways --- from Hawking's original calculation of quantum field theory in curved spacetime and Unruh's approach of an accelerated observer to Euclidean partition functions \cite{Bisognano:1976za, Gibbons:1976es, Gibbons:1978, Gibbons:1976ue}, canonical quantization in Hamiltonian formulation \cite{Carlip:1993sa}, quantum tunneling \cite{Parikh:1999mf,Vanzo:2011wq}, anomaly techniques \cite{Robinson:2005pd, Iso:2006wa, Iso:2006ut, Iso:2007nf, Bonora:2008nk, Bonora:2009tw}, the computation of pair production amplitudes \cite{Garfinkle:1993xk, Brown:1994um, Dowker:1993bt, Mann:1995vb}, and many others developed since early on to very recently --- that their existence seems very nearly certain!

The natural question is then whether black hole thermodynamics, like the ordinary thermodynamics of matter, has a microscopic ``statistical mechanical'' explanation. In 1996, using methods based on D-branes and string duality in string theory, Strominger and Vafa suggested a strategy of ``turning down'' the strength of the gravitational interaction until a black hole become a weakly coupled system of strings and D-branes so that at weak coupling one can count the number of states; this strategy led to the same Bekenstein-Hawking entropy for extremal (i.e., maximally charged) supersymmetric (BPS) black holes in five dimensions \cite{Strominger:1996sh}. This remarkable result was quickly extended to a large number of extremal, near-extremal, and some particular non-extremal black holes \cite{Peet:2000hn,Das:2000su}. However, it becomes harder and uncertain to obtain the right factor $\eta=1/4$ for far-from-extremal black holes (the Schwarzschild black hole is a typical case). This issue is currently under intense investigation in various approaches but has yet to be fully elucidated.

In 2005, Mathur proposed running the analysis backwards: starting at weak coupling with a particular collection of strings and D-branes, one then turns the gravitational coupling up and see what geometry appears at strong coupling \cite{Mathur:2005zp}. The result is typically not a black hole but instead a ``fuzzball'' --- a configuration with no horizon and no singularity but with a geometry akin to that of a black hole outside a would-be horizon \cite{Mathur:2005ai}. In a few special cases, one can count the number of such classical fuzzball geometries and reproduce the Bekenstein-Hawking entropy. In general, however, many of the relevant states may not have classical geometric descriptions.

In the context of string theory, the notable AdS/CFT correspondence also provides a feasible scheme to derive the black hole entropy. For asymptotically anti-de Sitter black holes, one can in principle compute the entropy by counting states in the (nongravitational) dual conformal field theory. The most straightforward application of this correspondence is for the (2+1)-dimensional BTZ black hole \cite{Banados:1992wn}. In 1997-1998, Strominger \cite{Strominger:1997eq} and Birmingham, Sachs, and Sen \cite{Birmingham:1998jt} independently computed the BTZ black hole entropy, which precisely reproduces the Bekenstein-Hawking expression. Since many higher dimensional near-extremal black holes have a near-horizon geometry of the form $\textit{BTZ}\times\textit{trivial}$, the BTZ results can be used to obtain the entropy of a large class of string theoretical black holes \cite{Skenderis:1999bs}.

The second major research program to derive the black hole entropy from a microscopic picture is loop quantum gravity. The suggestion that counting the quanta of area in loop quantum gravity could offer the statistical description of black hole thermodynamics was first proposed by Krasnov in 1997 \cite{Krasnov:1996tb}. The idea essentially is that the statistical ensemble is composed of the microstates of the horizon geometry that give rise to a specified total area. By introducing the concept of ``isolated horizons'',  Ashtekar \textit{et al.}\ rigorously developed the framework of the idea and derived the right Bekenstein-Hawking formula for the Schwarzschild black hole \cite{Ashtekar:1997yu, Ashtekar:1999wa, Ashtekar:2000eq}. These results were soon extended to other black holes with rotation, distortion, etc.\ \cite{Ashtekar:2000hw, Ashtekar:2001is, Ashtekar:2001jb, Ashtekar:2003jh, Ashtekar:2003zx, Ashtekar:2004nd}.

In 2004, the combinatorial problem of counting the microscopic states in loop quantum gravity was rephrased in a more manageable way by Domaga{\l}a and Lewandowski \cite{Domagala:2004jt} and by Meissner \cite{Meissner:2004ju}. These works corrected an flawed assumption considered true for several years and enabled one to compute the formula for $A_\mathrm{hor}\gg G\hbar$ to the subleading order as $k_\mathrm{B}^{-1}S = \frac{\gamma_\mathrm{M}}{\gamma}\frac{A_\mathrm{hor}}{4G\hbar}
-\frac{1}{2}\ln\frac{A}{G\hbar}+\cdots$, by which the Barbero-Immirzi parameter $\gamma$ is fixed as $\gamma=\gamma_\mathrm{M}\approx 0.23653$.
For small areas, the precise number counting was first suggested in 2006 \cite{Corichi:2006bs,Corichi:2006wn} and later thoroughly investigated by employing combinatorial methods (see Ref.~\cite{Agullo:2010zz} for a detailed account and Ref.~\cite{BarberoG.:2012ae} for a review). The key discovery is that, for microscopic black holes, the so-called black hole degeneracy spectrum when plotted as a function of the area exhibits a striking ``staircase'' structure, which makes contact in a nontrivial way with the evenly-spaced black hole horizon area spectrum predicted by Bekenstein and Mukhanov \cite{Bekenstein:1995ju}.
Recently, an explicit $SU(2)$ formulation for the black hole entropy has been developed by Engle \textit{et al.}\ based on covariant Hamiltonian methods \cite{Engle:2009vc,Engle:2010kt,Engle:2011vf}, giving rigorous support for the earlier proposal that the quantum black hole degrees of freedom could be described by an $SU(2)$ Chern-Simons theory. (Also see the review Ref.~\cite{BarberoG.:2012ae} and references therein for other recent advances.)

In addition to string theory and loop quantum gravity, we now have a number of statistical mechanical approaches to black hole thermodynamics. These include entanglement entropy \cite{Sorkin:2014kta, Bombelli:1986rw, Srednicki:1993im, Ryu:2006bv, Hubeny:2007xt}, induced gravity \cite{Sakharov:1967pk, Frolov:1996aj, Frolov:1997up}, and many others. While they may differ on the subleading corrections \cite{Carlip:2000nv}, they all seem to give the correct Bekenstein-Hawking entropy. (In general, one expects $k_\mathrm{B}^{-1}S = \frac{A_\mathrm{hor}}{4G\hbar} -\alpha\ln\frac{A}{G\hbar}+\cdots$, where the coefficient $\alpha$ depends on the quantum theory.) A new puzzle is why such a diverse set of quantum gravitational approaches should all agree. (There are some indications
that even the subleading corrections might be universal, up to differences on the treatment
of angular momentum and conserved charges \cite{Carlip:2000nv}.)

Even more puzzling is the ``information loss problem'': if a black hole is formed by collapse of matter in a pure quantum state, how can the seemingly thermal final state be compatible with unitary evolution?  This question is currently the subject of intense research, and may yield surprising new insights into the nature of space and time. (For a detailed account of the information paradox, see Ref.~\cite{Mathur:2008wi}.)

For a recent review of black hole thermodynamics, see Ref.~\cite{Carlip:2014pma}.

\section{Quantum gravity phenomenology}
Quantum gravity is often denounced as merely a theoretical enterprise that have no direct contact with experimental or observational realms, as quantum gravitational effects are appreciable only at the Planck scale, which is thought to be completely out of current reach. Today, as idea flourishes and technology advances drastically, the situation could be changed soon and quantum gravity might become reachable in the foreseeable future.

Many quantum gravity models suggest departures from the equivalence principle, CPT symmetry, and/or local Lorentz invariance in the deep Planck regime; among them the specific examples are string theory \cite{Kostelecky:1988zi}, brane-world scenarios \cite{Burgess:2002tb}, loop quantum gravity and spin foam theory \cite{Gambini:1998it,Rovelli:2010ed}, noncommutative geometry \cite{Lukierski:1993wx,AmelinoCamelia:1999pm,Carroll:2001ws}, emergent gravity \cite{Barcelo:2005fc}, etc. Deviations from the symmetries at the Planck scale in turn modify the pure metric spacetime structure and the local Lorentzian energy-momentum dispersion relation at lower-energy scale.

Modifications of the pure metric spacetime structure implies deviations from the Einstein Equivalence Principle (EEP). The most stringent test of EEP comes from cosmic observations. The null result of birefringence in cosmic propagation of polarized photons and polarized gamma rays so far has ascertained the light-cone structure and the core metric to ultrahigh precision of $10^{-38}$; this high precision already probes the second order in the ratio of $W$-boson mass or proton mass to the Planck mass \cite{Ni:2014qfa}. To this precision, the only freedoms left over regarding the metric structure are a scalar degree of freedom (dilaton) and a pseudo-scalar degree of freedom (axion). The dilaton alters the amplitude while the axion rotates the linear polarization of the cosmic propagation. Based on the cosmic propagation since the last scattering surface of the Cosmic Microwave Background (CMB), the fractional variation of the dilaton degree of freedom is constrained to $8\times10^{-4}$ by the precision agreement of the CMB spectrum to the black-body spectrum \cite{Ni:2014cca}. The cosmic axion degree of freedom is constrained by comic polarization rotation (CPR): uniform CPR to less than 0.02 (rad) \cite{Alighieri:2015hta}; CPR fluctuations to 0.02 (rad) \cite{Alighieri:2015hta,Mei:2014iaa}.

Observations on Ultrahigh Energy Cosmic Rays (UHECR) \cite{Kotera:2011cp} --- cosmic rays with kinetic energy greater than $10^{18}$ eV, the most energetic particles ever observed --- provide a chance to see possible modifications to the local Lorentzian energy-momentum dispersion relation of ultrarelativistic ($v\sim c$) particles. The UHECR spectrum measured at the Pierre Auger Observatory thus far has put very strong constraints on deviations from the Lorentzian dispersion relation \cite{Yamamoto:2007xj}. Future investigation with improved precision will either impose even stronger constraints or pick up signals of new physics including quantum gravitational effects (see Refs.~\cite{Stecker:2009hj,Liberati:2011bp} for detailed surveys).

Similarly, observations on propagation of ultrahigh energy cosmic electromagnetic waves can be used to test any momentum dependence of the speed of photons, as an indication of a breakdown of the Lorentz symmetry. Particularly, the detailed analysis of the relationship between the arrival time, photon momentum, and redshift of gamma-ray bursts is approaching very closely to the desired sensitivity to the Planck-scale physics \cite{AmelinoCamelia:2010zf}.

On the other hand, the ultra-precise measurements of ``atom-recoil frequency'' in the cold-atom experiments allow us to probe the energy-momentum dispersion relation of nonrelativistic ($v\ll c$) particles with Planck-scale sensitivity, thus providing revealing insight into various types of modifications to the dispersion relation considered in the quantum gravity literature \cite{AmelinoCamelia:2009zzb,Mercati:2010au}.
Combining studies of the two complementary regimes of the nonrelativistic and ultrarelativistic dispersion relations would tell us a great deal about the nature of quantum spacetime \cite{Mercati:2010au}.

In the context of astrophysics, many theories --- such as attractor theory \cite{Damour:1992kf}, the DGP model \cite{Dvali:2000hr}, massive gravity \cite{deRham:2014zqa}, etc.\ --- predict effects of quantum gravity in the solar system dynamics. None of these effects have been observed yet, but they might be testable in the proposed space missions designed to test relativistic gravity \cite{Ni:2008bj,Braxmaier:2011ai} and/or to detect and measure gravitational waves by using laser interferometry (see Ref.~\cite{Gair:2012nm} for a detailed survey on space-based gravitational-wave detectors).
Sensitive tests of the equivalence principle, local Lorentz invariance, modifications to
Einstein's field equations, etc.\ might also be possible in the coming space missions \cite{Lammerzahl:2004tc}.

In the context of cosmology, quantum effects in the pre-inflationary era may give rise to sufficient deviations from the standard inflationary scenario and leave footprints on the CMB.
In string theory, various observable imprints on anisotropies and polarization of the CMB have been suggested by different models of string cosmology and by the existence of cosmic strings (see Ref.~\cite{Baumann:2014nda} for a review).
In loop quantum gravity, the standard inflationary scenario is extended to the epoch from the big bounce in the Planck era to the onset of slow-roll inflation \cite{Ashtekar:2013xka}. Some research works have attempted to reveal possible observable imprints of loop quantum effects on the CMB \cite{Grain:2009kw, Barrau:2009fz, Mielczarek:2010bh, Grain:2010yv, Barrau:2010nd, Agullo:2013ai}.

There are a few experiments currently in operation or in ongoing development, aiming to make accurate measurements of anisotropies and polarization of the CMB, and their results will soon lead to great excitement. Very recently, the joint analysis of BICEP2/\textit{Keck Array} and \textit{Planck} Data found no statistically significant evidence for tensor modes in the $B$-mode polarization, yielding an upper limit $r<0.12$ at 95\% confidence on the tensor-to-scalar ratio $r$ \cite{Ade:2015tva}. This upper limit already ruled out many ``large-$r$ scenarios'' of pre-inflationary models, including those in string cosmology. The conclusive result of $B$-mode polarization expected to arrive in the near future will further discriminate between different quantum gravity theories.

There are many other cosmological and astrophysical observations as well as laboratory experiments that could reveal possible quantum gravitational effects. At the turn of the centennial of the birth of Einstein's general relativity, quantum gravity phenomenology will in time become an important area of relevant research. (For a comprehensive survey on quantum gravity phenomenology, see Ref.~\cite{AmelinoCamelia:2008qg}; also see \cite{Smolin:2005re} for a survey from the perspective of loop quantum gravity.)




\begin{thebibliography}{99}


\subsection*{Prelude}


\bibitem{Einstein:1916}
  A.~Einstein,
  ``Noherungsweise Integration der Feldgleichungen der Gravitation,''
  Preussische Akademie der Wissenschaften (Berlin). Sitzungsberichte 688-696 (1916).

\bibitem{Klein:1927}
  O.~Klein,
  ``Zur Fuenfdimentionalen Darstellung der relativitaetstheorie,''
  A.\ fuer Physik {\bf 46}, 688 (1927).

\bibitem{Klein:1954}
  O.~Klein,
  ``Aktuella Problem Kring Fysikens Sma och Stora Tal,''
  Kosmos (Sweden) {\bf 32} 33 (1954).

\bibitem{Klein:1956}
  O.~Klein,
  ``Generalization of Einstein’s theory of gravitation considered from the point of view of quantum field theory,''
  in \emph{Fuenfzig Jahre Relativit\"{a}tstheorie.\ Bern, 11--16 Juli 1955}
  (Helvetica Physica Acta, Suppl.\ 4, eds.\ A.~Mercier and M.~Kervaire)
  (Basel: Birkh\"{a}user, 1956).

\bibitem{Deser:1957zz}
  S.~Deser,
  ``General Relativity and the Divergence Problem in Quantum Field Theory,''
  Rev.\ Mod.\ Phys.\  {\bf 29}, 417 (1957).

\bibitem{Stachel:1916}
 J.~Stachel, ``Early history of quantum gravity (1916-1940),'' in \emph{Black Holes, Gravitational Radiation and the Universe}, eds.\ R.~Iyer and B.~Bhawal (Kluwer, 1999).

\bibitem{Fierz:1939ix}
  M.~Fierz and W.~Pauli,
  ``On relativistic wave equations for particles of arbitrary spin in an electromagnetic field,''
  Proc.\ Roy.\ Soc.\ Lond.\ A {\bf 173}, 211 (1939).

\bibitem{Deser:1959zza}
  R.~Arnowitt and S.~Deser,
  ``Quantum Theory of Gravitation: General Formulation and Linearized Theory,''
  Phys.\ Rev.\  {\bf 113}, 745 (1959).

\bibitem{Arnowitt:1959ah}
  R.~L.~Arnowitt, S.~Deser and C.~W.~Misner,
  ``Dynamical Structure and Definition of Energy in General Relativity,''
  Phys.\ Rev.\  {\bf 116}, 1322 (1959).
\bibitem{Arnowitt:1960es}
  R.~L.~Arnowitt, S.~Deser and C.~W.~Misner,
  ``Canonical variables for general relativity,''
  Phys.\ Rev.\  {\bf 117}, 1595 (1960).

\bibitem{Deser:1960zza}
  S.~Deser, R.~Arnowitt and C.~W.~Misner,
  ``Canonical Variables, Energy and Criteria for Radiation in General Relativity,''
  Nuovo Cim.\  {\bf 15}, 487 (1960).

\bibitem{Arnowitt:1960zzc}
  R.~Arnowitt, S.~Deser and C.~W.~Misner,
  ``Energy and the Criteria for Radiation in General Relativity,''
  Phys.\ Rev.\  {\bf 118}, 1100 (1960).

\bibitem{Deser:1960zzc}
  S.~Deser, R.~Arnowitt and C.~W.~Misner,
  ``Consistency of Canonical Reduction of General Relativity,''
  J.\ Math.\ Phys.\  {\bf 1}, 434 (1960).

\bibitem{Arnowitt:1960zza}
  R.~L.~Arnowitt, S.~Deser and C.~W.~Misner,
  ``Gravitational-electromagnetic coupling and the classical self-energy problem,''
  Phys.\ Rev.\  {\bf 120}, 313 (1960).

\bibitem{Deser:1960zzb}
  S.~Deser, R.~Arnowitt and C.~W.~Misner,
  ``Note on Positive-Definiteness of Energy of the Gravitational Field,''
  Annals Phys.\  {\bf 11}, 116 (1960).

\bibitem{Deser:1961zza}
  S.~Deser, R.~Arnowitt and C.~W.~Misner,
  ``Heisenberg Representation in Classical General Relativity,''
  Nuovo Cim.\  {\bf 19}, 668 (1961).

\bibitem{Arnowitt:1961zza}
  R.~L.~Arnowitt, S.~Deser and C.~W.~Misner,
  ``Wave zone in general relativity,''
  Phys.\ Rev.\  {\bf 121}, 1556 (1961).

\bibitem{Arnowitt:1961zz}
  R.~L.~Arnowitt, S.~Deser and C.~W.~Misner,
  ``Coordinate invariance and energy expressions in general relativity,''
  Phys.\ Rev.\  {\bf 122}, 997 (1961).

\bibitem{Arnowitt:1962hi}
  R.~L.~Arnowitt, S.~Deser and C.~W.~Misner,
  ``Republication of: The Dynamics of general relativity,''
  Gen.\ Rel.\ Grav.\  {\bf 40}, 1997 (2008);
  originally appeared as Chap.~7, pp.~227--264, in \emph{Gravitation: an introduction to current research}, ed.\ L.~Witten (Wiley, New York, 1962)
  [gr-qc/0405109].

\bibitem{Schoen:1979rg}
  R.~Schoen and S.~T.~Yau,
  ``On the Proof of the positive mass conjecture in general relativity,''
  Commun.\ Math.\ Phys.\  {\bf 65}, 45 (1979).

\bibitem{Schoen:1981vd}
  R.~Schoen and S.~T.~Yau,
  ``Proof of the positive mass theorem. 2.,''
  Commun.\ Math.\ Phys.\  {\bf 79}, 231 (1981).



\bibitem{Rovelli:1997qj}
  C.~Rovelli,
  ``Strings, loops and others: A Critical survey of the present approaches to quantum gravity,''
  in \emph{Gravitation and Relativity: At the Turn of the Millenium}
  (Inter-University Centre for Astronomy and Astrophysics, Pune, 1998), pp.\ 281--331
  [gr-qc/9803024].

\bibitem{Carlip:2001wq}
  S.~Carlip,
  ``Quantum gravity: A Progress report,''
  Rept.\ Prog.\ Phys.\  {\bf 64}, 885, (2001)
  [gr-qc/0108040].

\bibitem{Woodard:2009ns}
  R.~P.~Woodard,
  ``How Far Are We from the Quantum Theory of Gravity?,''
  Rept.\ Prog.\ Phys.\  {\bf 72}, 126002 (2009)
  [arXiv:0907.4238 [gr-qc]].

\bibitem{Kiefer:2012boa}
  C.~Kiefer,
  \emph{Quantum Gravity},
  (Oxford University Press, Oxford, 2012).


\bibitem{Rovelli:2000aw}
  C.~Rovelli,
  ``Notes for a brief history of quantum gravity,''
  gr-qc/0006061.


\bibitem{Smolin:2000af}
  L.~Smolin,
  \emph{Three Roads to Quantum Gravity},
  (Weidenfeld and Nicolson, London, 2000).


\subsection*{Perturbative quantum gravity}

\bibitem{DeWitt:1967yk}
  B.~S.~DeWitt,
  ``Quantum Theory of Gravity. 1. The Canonical Theory,''
  Phys.\ Rev.\  {\bf 160}, 1113 (1967).

\bibitem{DeWitt:1967ub}
  B.~S.~DeWitt,
  ``Quantum Theory of Gravity. 2. The Manifestly Covariant Theory,''
  Phys.\ Rev.\  {\bf 162}, 1195 (1967).

\bibitem{DeWitt:1967uc}
  B.~S.~DeWitt,
  ``Quantum Theory of Gravity. 3. Applications of the Covariant Theory,''
  Phys.\ Rev.\  {\bf 162}, 1239 (1967).


\bibitem{Abbott:1981ke}
  L.~F.~Abbott,
  ``Introduction to the Background Field Method,''
  Acta Phys.\ Polon.\ B {\bf 13}, 33 (1982).

\bibitem{Faddeev:1967fc}
  L.~D.~Faddeev and V.~N.~Popov,
  ``Feynman Diagrams for the Yang-Mills Field,''
  Phys.\ Lett.\ B {\bf 25}, 29 (1967).

\bibitem{'tHooft:1972fi}
  G.~'t Hooft and M.~J.~G.~Veltman,
  ``Regularization and Renormalization of Gauge Fields,''
  Nucl.\ Phys.\ B {\bf 44}, 189 (1972).

\bibitem{Bollini:1972ui}
  C.~G.~Bollini and J.~J.~Giambiagi,
  ``Dimensional Renormalization: The Number of Dimensions as a Regularizing Parameter,''
  Nuovo Cim.\  {\bf 12}, no. 1, 20 (1972).

\bibitem{'tHooft:1974bx}
  G.~'t Hooft and M.~J.~G.~Veltman,
  ``One loop divergencies in the theory of gravitation,''
  Annales Poincare Phys.\ Theor.\ A {\bf 20}, 69 (1974).

\bibitem{Deser:1974zzd}
  S.~Deser and P.~van Nieuwenhuizen,
  ``Nonrenormalizability of the Quantized Einstein-Maxwell System,''
  Phys.\ Rev.\ Lett.\  {\bf 32}, 245 (1974).

\bibitem{Deser:1974cz}
  S.~Deser and P.~van Nieuwenhuizen,
  ``One Loop Divergences of Quantized Einstein-Maxwell Fields,''
  Phys.\ Rev.\ D {\bf 10}, 401 (1974).

\bibitem{Deser:1974cy}
  S.~Deser and P.~van Nieuwenhuizen,
  ``Nonrenormalizability of the Quantized Dirac-Einstein System,''
  Phys.\ Rev.\ D {\bf 10}, 411 (1974).

\bibitem{Deser:1974xq}
  S.~Deser, H.~S.~Tsao and P.~van Nieuwenhuizen,
 `` One Loop Divergences of the Einstein Yang-Mills System,''
  Phys.\ Rev.\ D {\bf 10}, 3337 (1974).

\bibitem{Goroff:1985sz}
  M.~H.~Goroff and A.~Sagnotti,
  ``Quantum Gravity At Two Loops,''
  Phys.\ Lett.\ B {\bf 160}, 81 (1985).

\bibitem{Goroff:1985th}
  M.~H.~Goroff and A.~Sagnotti,
  ``The Ultraviolet Behavior of Einstein Gravity,''
  Nucl.\ Phys.\ B {\bf 266}, 709 (1986).

\bibitem{vandeVen:1991gw}
  A.~E.~M.~van de Ven,
  ``Two loop quantum gravity,''
  Nucl.\ Phys.\ B {\bf 378}, 309 (1992).

\bibitem{Donoghue:1993eb}
  J.~F.~Donoghue,
  ``Leading quantum correction to the Newtonian potential,''
  Phys.\ Rev.\ Lett.\  {\bf 72}, 2996 (1994)
  [gr-qc/9310024].

\bibitem{Donoghue:1994dn}
  J.~F.~Donoghue,
  ``General relativity as an effective field theory: The leading quantum corrections,''
  Phys.\ Rev.\ D {\bf 50}, 3874 (1994)
  [gr-qc/9405057].

\bibitem{Feinberg:1968zz}
  G.~Feinberg and J.~Sucher,
  ``Long-Range Forces from Neutrino-Pair Exchange,''
  Phys.\ Rev.\  {\bf 166}, 1638 (1968).

\bibitem{Hsu:1992tg}
  S.~D.~H.~Hsu and P.~Sikivie,
  ``Long range forces from two neutrino exchange revisited,''
  Phys.\ Rev.\ D {\bf 49}, 4951 (1994)
  [hep-ph/9211301].

\bibitem{DeWitt:1975ys}
  B.~S.~DeWitt,
  ``Quantum Field Theory in Curved Space-Time,''
  Phys.\ Rept.\  {\bf 19}, 295 (1975).

\bibitem{Schwinger:1960qe}
  J.~S.~Schwinger,
  ``Brownian motion of a quantum oscillator,''
  J.\ Math.\ Phys.\  {\bf 2}, 407 (1961).

\bibitem{Mahanthappa:1962ex}
  K.~T.~Mahanthappa,
  ``Multiple production of photons in quantum electrodynamics,''
  Phys.\ Rev.\  {\bf 126}, 329 (1962).

\bibitem{Bakshi:1962dv}
  P.~M.~Bakshi and K.~T.~Mahanthappa,
  ``Expectation value formalism in quantum field theory. 1.,''
  J.\ Math.\ Phys.\  {\bf 4}, 1 (1963).

\bibitem{Bakshi:1963bn}
  P.~M.~Bakshi and K.~T.~Mahanthappa,
  ``Expectation value formalism in quantum field theory. 2.,''
  J.\ Math.\ Phys.\  {\bf 4}, 12 (1963).

\bibitem{Keldysh:1964ud}
  L.~V.~Keldysh,
  ``Diagram technique for nonequilibrium processes,''
  Zh.\ Eksp.\ Teor.\ Fiz.\  {\bf 47}, 1515 (1964)
  [Sov.\ Phys.\ JETP {\bf 20}, 1018 (1965)].

\bibitem{Chou:1984es}
  K.~c.~Chou, Z.~b.~Su, B.~l.~Hao and L.~Yu,
  ``Equilibrium and Nonequilibrium Formalisms Made Unified,''
  Phys.\ Rept.\  {\bf 118}, 1 (1985).

\bibitem{Jordan:1986ug}
  R.~D.~Jordan,
  ``Effective Field Equations for Expectation Values,''
  Phys.\ Rev.\ D {\bf 33}, 444 (1986).

\bibitem{Calzetta:1986ey}
  E.~Calzetta and B.~L.~Hu,
  ``Closed Time Path Functional Formalism in Curved Space-Time: Application to Cosmological Back Reaction Problems,''
  Phys.\ Rev.\ D {\bf 35}, 495 (1987).

\bibitem{Calzetta:1986cq}
  E.~Calzetta and B.~L.~Hu,
  ``Nonequilibrium Quantum Fields: Closed Time Path Effective Action, Wigner Function and Boltzmann Equation,''
  Phys.\ Rev.\ D {\bf 37}, 2878 (1988).

\bibitem{Weinberg:2005vy}
  S.~Weinberg,
  ``Quantum contributions to cosmological correlations,''
  Phys.\ Rev.\ D {\bf 72}, 043514 (2005)
  [hep-th/0506236].

\bibitem{Weinberg:2006ac}
  S.~Weinberg,
  ``Quantum contributions to cosmological correlations. II. Can these corrections become large?,''
  Phys.\ Rev.\ D {\bf 74}, 023508 (2006)
  [hep-th/0605244].


\bibitem{Woodard:2014jba}
  R.~P.~Woodard,
  ``Perturbative Quantum Gravity Comes of Age,''
  Int.\ J.\ Mod.\ Phys.\ D {\bf 23}, no.\ 09, 1430020 (2014)
  [arXiv:1407.4748 [gr-qc]].


\subsection*{String theory}

\bibitem{Green:1987sp}
  M.~B.~Green, J.~H.~Schwarz and E.~Witten,
  \emph{Superstring Theory. Vol.\ 1: Introduction, Vol.\ 2: Loop Amplitudes, Anomalies And Phenomenology},
  (Cambridge University Press, Cambridge, 1987).

\bibitem{Polchinski:1998rq}
  J.~Polchinski,
  \emph{String Theory. Vol.\ 1: An Introduction to the Bosonic String, Vol.\ 2: Superstring Theory and Beyond},
  (Cambridge University Press, Cambridge, 1998).

\bibitem{Becker:2007zj}
  K.~Becker, M.~Becker and J.~H.~Schwarz,
  \emph{String theory and M-theory: A modern introduction},
  (Cambridge University Press, Cambridge, 2007).

\bibitem{Rickles:2014fha}
  D.~Rickles,
  \emph{A brief history of string theory: From dual models to M-theory},
  (Springer-Verlag, Berlin Heidelberg, 2014).

\bibitem{Schwarz:2007yc}
  J.~H.~Schwarz,
  ``The Early Years of String Theory: A Personal Perspective,''
  arXiv:0708.1917 [hep-th].

\bibitem{Nambu:1986ze}
  Y.~Nambu,
  ``Duality And Hadrodynamics,''
  Notes prepared for the Copenhagen High Energy Symposium (1970);
  Reprinted in \emph{Broken symmetry: Selected Papers of Y.\ Nambu}, eds.\ T.~Eguchi and K.~Nishijima (Singopore, World Scientific, 1995).

\bibitem{Goto:1971ce}
  T.~Goto,
  ``Relativistic quantum mechanics of one-dimensional mechanical continuum and subsidiary condition of dual resonance model,''
  Prog.\ Theor.\ Phys.\  {\bf 46}, 1560 (1971).

\bibitem{Hara:1971ur}
  O.~Hara,
  ``On origin and physical meaning of ward-like identity in dual-resonance model,''
  Prog.\ Theor.\ Phys.\  {\bf 46}, 1549 (1971).

\bibitem{Polyakov:1981rd}
  A.~M.~Polyakov,
  ``Quantum Geometry of Bosonic Strings,''
  Phys.\ Lett.\ B {\bf 103}, 207 (1981).

\bibitem{Polyakov:1981re}
  A.~M.~Polyakov,
  ``Quantum Geometry of Fermionic Strings,''
  Phys.\ Lett.\ B {\bf 103}, 211 (1981).

\bibitem{Callan:1985ia}
  C.~G.~Callan, Jr., E.~J.~Martinec, M.~J.~Perry and D.~Friedan,
  ``Strings in Background Fields,''
  Nucl.\ Phys.\ B {\bf 262}, 593 (1985).

\bibitem{Ramond:1971gb}
  P.~Ramond,
  ``Dual Theory for Free Fermions,''
  Phys.\ Rev.\ D {\bf 3}, 2415 (1971).

\bibitem{Neveu:1971rx}
  A.~Neveu and J.~H.~Schwarz,
  ``Factorizable dual model of pions,''
  Nucl.\ Phys.\ B {\bf 31}, 86 (1971).

\bibitem{Green:1980zg}
  M.~B.~Green and J.~H.~Schwarz,
  ``Supersymmetrical Dual String Theory,''
  Nucl.\ Phys.\ B {\bf 181}, 502 (1981).

\bibitem{Green:1981xx}
  M.~B.~Green and J.~H.~Schwarz,
  ``Supersymmetrical Dual String Theory. 2. Vertices and Trees,''
  Nucl.\ Phys.\ B {\bf 198}, 252 (1982).

\bibitem{Green:1981yb}
  M.~B.~Green and J.~H.~Schwarz,
  ``Supersymmetrical String Theories,''
  Phys.\ Lett.\ B {\bf 109}, 444 (1982).

\bibitem{Green:1984sg}
  M.~B.~Green and J.~H.~Schwarz,
  ``Anomaly Cancellation in Supersymmetric D=10 Gauge Theory and Superstring Theory,''
  Phys.\ Lett.\ B {\bf 149}, 117 (1984).

\bibitem{Giveon:1994fu}
  A.~Giveon, M.~Porrati and E.~Rabinovici,
  ``Target space duality in string theory,''
  Phys.\ Rept.\  {\bf 244}, 77 (1994)
  [hep-th/9401139].

\bibitem{Alvarez:1994dn}
  E.~Alvarez, L.~Alvarez-Gaume and Y.~Lozano,
  ``An Introduction to T duality in string theory,''
  Nucl.\ Phys.\ Proc.\ Suppl.\  {\bf 41}, 1 (1995)
  [hep-th/9410237].

\bibitem{Polchinski:1995df}
  J.~Polchinski and E.~Witten,
  ``Evidence for heterotic - type I string duality,''
  Nucl.\ Phys.\ B {\bf 460}, 525 (1996)
  [hep-th/9510169].

\bibitem{Hull:1994ys}
  C.~M.~Hull and P.~K.~Townsend,
  ``Unity of superstring dualities,''
  Nucl.\ Phys.\ B {\bf 438}, 109 (1995)
  [hep-th/9410167].

\bibitem{Polchinski:1995mt}
  J.~Polchinski,
  ``Dirichlet Branes and Ramond-Ramond charges,''
  Phys.\ Rev.\ Lett.\  {\bf 75}, 4724 (1995)
  [hep-th/9510017].

\bibitem{Townsend:1995kk}
  P.~K.~Townsend,
  ``The eleven-dimensional supermembrane revisited,''
  Phys.\ Lett.\ B {\bf 350}, 184 (1995)
  [hep-th/9501068].

\bibitem{Witten:1995ex}
  E.~Witten,
  ``String theory dynamics in various dimensions,''
  Nucl.\ Phys.\ B {\bf 443}, 85 (1995)
  [hep-th/9503124].

\bibitem{Horava:1995qa}
  P.~Ho\v{r}ava and E.~Witten,
  ``Heterotic and type I string dynamics from eleven-dimensions,''
  Nucl.\ Phys.\ B {\bf 460}, 506 (1996)
  [hep-th/9510209].

\bibitem{Horava:1996ma}
  P.~Ho\v{r}ava and E.~Witten,
  ``Eleven-dimensional supergravity on a manifold with boundary,''
  Nucl.\ Phys.\ B {\bf 475}, 94 (1996)
  [hep-th/9603142].

\bibitem{Maldacena:1997re}
  J.~M.~Maldacena,
  ``The Large N limit of superconformal field theories and supergravity,''
  Int.\ J.\ Theor.\ Phys.\  {\bf 38}, 1113 (1999)
  [Adv.\ Theor.\ Math.\ Phys.\  {\bf 2}, 231 (1998)]
  [hep-th/9711200].

\bibitem{Aharony:1999ti}
  O.~Aharony, S.~S.~Gubser, J.~M.~Maldacena, H.~Ooguri and Y.~Oz,
  ``Large N field theories, string theory and gravity,''
  Phys.\ Rept.\  {\bf 323}, 183 (2000)
  [hep-th/9905111].

\bibitem{'tHooft:1993gx}
  G.~'t Hooft,
  ``Dimensional reduction in quantum gravity,''
  in \emph{Salamfestschrift: A Collection of Talks from the Conference on Highlights of Particle and Condensed Matter Physics, Trieste, Italy 8-12 March 1993}
  (World Scientific, 1993)
  [gr-qc/9310026].

\bibitem{Susskind:1994vu}
  L.~Susskind,
  ``The World as a hologram,''
  J.\ Math.\ Phys.\  {\bf 36}, 6377 (1995)
  [hep-th/9409089].

\bibitem{Strominger:1986uh}
  A.~Strominger,
  ``Superstrings with Torsion,''
  Nucl.\ Phys.\ B {\bf 274}, 253 (1986).

\bibitem{de Wit:1986xg}
  B.~de Wit, D.~J.~Smit and N.~D.~Hari Dass,
  ``Residual Supersymmetry of Compactified $D=10$ Supergravity,''
  Nucl.\ Phys.\ B {\bf 283}, 165 (1987).

\bibitem{Becker:1996gj}
  K.~Becker and M.~Becker,
  ``M theory on eight manifolds,''
  Nucl.\ Phys.\ B {\bf 477}, 155 (1996)
  [hep-th/9605053].

\bibitem{Grana:2005jc}
  M.~Gra\~{n}a,
  ``Flux compactifications in string theory: A Comprehensive review,''
  Phys.\ Rept.\  {\bf 423}, 91 (2006)
  [hep-th/0509003].

\bibitem{Gukov:1999ya}
  S.~Gukov, C.~Vafa and E.~Witten,
  ``CFT's from Calabi-Yau four-folds,''
  Nucl.\ Phys.\ B {\bf 584}, 69 (2000)
  [Erratum: \textit{ibid.}\ {\bf 608}, 477 (2001)]
  [hep-th/9906070].

\bibitem{Douglas:2003um}
  M.~R.~Douglas,
  ``The Statistics of string/M theory vacua,''
  JHEP {\bf 0305}, 046 (2003)
  [hep-th/0303194].

\bibitem{Susskind:2003kw}
  L.~Susskind,
  ``The Anthropic landscape of string theory,''
  in \emph{Universe or multiverse?} pp.~247-266,
  ed.\ C.~Bernard
  (Cambridge University Press, 2009)
  [hep-th/0302219].

\bibitem{Randall:1999vf}
  L.~Randall and R.~Sundrum,
  ``An Alternative to compactification,''
  Phys.\ Rev.\ Lett.\  {\bf 83}, 4690 (1999)
  [hep-th/9906064].

\bibitem{Kachru:2003aw}
  S.~Kachru, R.~Kallosh, A.~D.~Linde and S.~P.~Trivedi,
  ``De Sitter vacua in string theory,''
  Phys.\ Rev.\ D {\bf 68}, 046005 (2003)
  [hep-th/0301240].

\bibitem{Kachru:2003sx}
  S.~Kachru, R.~Kallosh, A.~D.~Linde, J.~M.~Maldacena, L.~P.~McAllister and S.~P.~Trivedi,
  ``Towards inflation in string theory,''
  JCAP {\bf 0310}, 013 (2003)
  [hep-th/0308055].

\bibitem{Quevedo:2002xw}
  F.~Quevedo,
  ``Lectures on string/brane cosmology,''
  Class.\ Quant.\ Grav.\  {\bf 19}, 5721 (2002)
  [hep-th/0210292].

\bibitem{Danielsson:2004mf}
  U.~H.~Danielsson,
  ``Lectures on string theory and cosmology,''
  Class.\ Quant.\ Grav.\  {\bf 22}, S1 (2005)
  [hep-th/0409274].

\bibitem{Baumann:2014nda}
  D.~Baumann and L.~McAllister,
  \emph{Inflation and String Theory},
  (Cambridge University Press, Cambridge, 2015)
  [arXiv:1404.2601 [hep-th]].


\subsection*{Loop quantum gravity}





\bibitem{Rovelli:2004tv}
  C.~Rovelli,
  \emph{Quantum Gravity},
  (Cambridge University Press, Cambridge, 2004).

\bibitem{Thiemann:2007zz}
  T.~Thiemann,
  \emph{Modern Canonical Quantum General Relativity},
  (Cambridge University Press, Cambridge, 2007)
  [gr-qc/0110034].



\bibitem{Rovelli:2014book}
 C.~Rovelli and F.~Vidotto
 \emph{Covariant Loop Quantum Gravity: An Elementary Introduction to Quantum Gravity and Spinfoam Theory},
 (Cambridge University Press, Cambridge, 2014.)


\bibitem{Rovelli:2010bf}
  C.~Rovelli,
  "Loop quantum gravity: the first twenty five years,"
  Class.\ Quant.\ Grav.\  {\bf 28}, 153002, (2011)
  [arXiv:1012.4707 [gr-qc]].


\bibitem{Chiou:2014jwa}
  D.~W.~Chiou,
  ``Loop Quantum Gravity,''
  Int.\ J.\ Mod.\ Phys.\ D {\bf 24}, no.\ 01, 1530005 (2014)
  [arXiv:1412.4362 [gr-qc]].


\bibitem{Ashtekar:1986yd}
  A.~Ashtekar,
 `` New Variables for Classical and Quantum Gravity,''
  Phys.\ Rev.\ Lett.\  {\bf 57}, 2244 (1986).

\bibitem{Ashtekar:1987gu}
  A.~Ashtekar,
  ``New Hamiltonian Formulation of General Relativity,''
  Phys.\ Rev.\ D {\bf 36}, 1587 (1987).


\bibitem{Jacobson:1987qk}
  T.~Jacobson and L.~Smolin,
  ``Nonperturbative Quantum Geometries,''
  Nucl.\ Phys.\ B {\bf 299}, 295 (1988).

\bibitem{Rovelli:1987df}
  C.~Rovelli and L.~Smolin,
  ``Knot Theory and Quantum Gravity,''
  Phys.\ Rev.\ Lett.\  {\bf 61}, 1155 (1988).

\bibitem{Rovelli:1989za}
  C.~Rovelli and L.~Smolin,
  ``Loop Space Representation of Quantum General Relativity,''
  Nucl.\ Phys.\ B {\bf 331}, 80 (1990).

\bibitem{Rovelli:1991zi}
  C.~Rovelli,
  ``Ashtekar formulation of general relativity and loop space nonperturbative quantum gravity: A Report,''
  Class.\ Quant.\ Grav.\  {\bf 8}, 1613 (1991).

\bibitem{Bruegmann:1992ak}
  B.~Br\"{u}gmann, R.~Gambini and J.~Pullin,
  ``Knot invariants as nondegenerate quantum geometries,''
  Phys.\ Rev.\ Lett.\  {\bf 68}, 431 (1992).

\bibitem{Bruegmann:1992gp}
  B.~Br\"{u}gmann, R.~Gambini and J.~Pullin,
  ``Jones polynomials for intersecting knots as physical states of quantum gravity,''
  Nucl.\ Phys.\ B {\bf 385}, 587 (1992)
  [hep-th/9202018].

\bibitem{Penrose:1971}
  R.~Penrose,
  ``Angular momentum: an approach to combinatorial spacetime,''
  in \emph{Quantum Theory and Beyond}, ed.\ T.~Bastin (Cambridge University Press, Cambridge 1971).

\bibitem{Rovelli:1995ac}
  C.~Rovelli and L.~Smolin,
  ``Spin networks and quantum gravity,''
  Phys.\ Rev.\ D {\bf 52}, 5743 (1995)
  [gr-qc/9505006].

\bibitem{Baez:1994hx}
  J.~C.~Baez,
  ``Spin network states in gauge theory,''
  Adv.\ Math.\  {\bf 117}, 253 (1996)
  [gr-qc/9411007].

\bibitem{Baez:1995md}
  J.~C.~Baez,
  ``Spin networks in nonperturbative quantum gravity,''
  In *San Francisco 1995, The interface of knots and physics* 167-203
  [gr-qc/9504036].

\bibitem{Ashtekar:1992tm}
  A.~Ashtekar, C.~Rovelli and L.~Smolin,
  ``Weaving a classical geometry with quantum threads,''
  Phys.\ Rev.\ Lett.\  {\bf 69}, 237 (1992)
  [hep-th/9203079].

\bibitem{Rovelli:1994ge}
  C.~Rovelli and L.~Smolin,
  ``Discreteness of area and volume in quantum gravity,''
  Nucl.\ Phys.\ B {\bf 442}, 593 (1995)
  [Erratum: Nucl.\ Phys.\ B {\bf 456}, 753 (1995)]
  [gr-qc/9411005].

\bibitem{De Pietri:1996pja}
  R.~De Pietri and C.~Rovelli,
  ``Geometry eigenvalues and scalar product from recoupling theory in loop quantum gravity,''
  Phys.\ Rev.\ D {\bf 54}, 2664 (1996)
  [gr-qc/9602023].

\bibitem{Thiemann:1996au}
  T.~Thiemann,
  ``Closed formula for the matrix elements of the volume operator in canonical quantum gravity,''
  J.\ Math.\ Phys.\  {\bf 39}, 3347 (1998)
  [gr-qc/9606091].

\bibitem{Lewandowski:1996gk}
  J.~Lewandowski,
  ``Volume and quantizations,''
  Class.\ Quant.\ Grav.\  {\bf 14}, 71 (1997)
  [gr-qc/9602035].


\bibitem{Thiemann:1996ay}
  T.~Thiemann,
  ``Anomaly-free formulation of nonperturbative, four-dimensional Lorentzian quantum gravity,''
  Phys.\ Lett.\ B {\bf 380}, 257 (1996)
  [gr-qc/9606088].

\bibitem{Thiemann:1996aw}
  T.~Thiemann,
  ``Quantum spin dynamics (QSD),''
  Class.\ Quant.\ Grav.\  {\bf 15}, 839 (1998)
  [gr-qc/9606089].

\bibitem{Thiemann:1996av}
  T.~Thiemann,
  ``Quantum spin dynamics (QSD): II.\ The kernel of the Wheeler-DeWitt constraint operator,''
  Class.\ Quant.\ Grav.\  {\bf 15}, 875 (1998)
  [gr-qc/9606090].

\bibitem{Thiemann:1997rv}
  T.~Thiemann,
  ``Quantum spin dynamics (QSD): III.\ Quantum constraint algebra and physical scalar product in quantum general relativity,''
  Class.\ Quant.\ Grav.\  {\bf 15}, 1207 (1998)
  [gr-qc/9705017].


\bibitem{Thiemann:2003zv}
  T.~Thiemann,
  ``The Phoenix project: Master constraint program for loop quantum gravity,''
  Class.\ Quant.\ Grav.\  {\bf 23}, 2211 (2006)
  [gr-qc/0305080].

\bibitem{Giesel:2006uj}
  K.~Giesel and T.~Thiemann,
  Algebraic Quantum Gravity (AQG): I.\ Conceptual Setup,
  \emph{Class.\ Quant.\ Grav.}\  {\bf 24}, 2465 (2007)
  [gr-qc/0607099].

\bibitem{Giesel:2006uk}
  K.~Giesel and T.~Thiemann,
  Algebraic Quantum Gravity (AQG): II.\ Semiclassical Analysis,
  \emph{Class.\ Quant.\ Grav.}\  {\bf 24}, 2499 (2007)
  [gr-qc/0607100].

\bibitem{Giesel:2006um}
  K.~Giesel and T.~Thiemann,
  ``Algebraic quantum gravity (AQG): III.\ Semiclassical perturbation theory,''
  Class.\ Quant.\ Grav.\  {\bf 24}, 2565 (2007)
  [gr-qc/0607101].

\bibitem{Giesel:2007wn}
  K.~Giesel and T.~Thiemann,
  ``Algebraic quantum gravity (AQG): IV.\ Reduced phase space quantisation of loop quantum gravity,''
  Class.\ Quant.\ Grav.\  {\bf 27}, 175009 (2010)
  [arXiv:0711.0119 [gr-qc]].


\bibitem{Ashtekar:1994nx}
  A.~Ashtekar, J.~Lewandowski, D.~Marolf, J.~Mourao and T.~Thiemann,
  ``Coherent state transforms for spaces of connections,''
  J.\ Funct.\ Anal.\  {\bf 135}, 519 (1996)
  [gr-qc/9412014].

\bibitem{Thiemann:2002vj}
  T.~Thiemann,
  ``Complexifier coherent states for quantum general relativity,''
  Class.\ Quant.\ Grav.\  {\bf 23}, 2063 (2006)
  [gr-qc/0206037].

\bibitem{Bahr:2007xa}
  B.~Bahr and T.~Thiemann,
  ``Gauge-invariant coherent states for Loop Quantum Gravity: I.\ Abelian gauge groups,''
  Class.\ Quant.\ Grav.\  {\bf 26}, 045011 (2009)
  [arXiv:0709.4619 [gr-qc]].

\bibitem{Bahr:2007xn}
  B.~Bahr and T.~Thiemann,
  ``Gauge-invariant coherent states for loop quantum gravity: II.\ Non-Abelian gauge groups,''
  Class.\ Quant.\ Grav.\  {\bf 26}, 045012 (2009)
  [arXiv:0709.4636 [gr-qc]].

\bibitem{Flori:2009rw}
  C.~Flori,
  ``Semiclassical analysis of the Loop Quantum Gravity volume operator: Area Coherent States,''
  arXiv:0904.1303 [gr-qc].

\bibitem{Bianchi:2009ky}
  E.~Bianchi, E.~Magliaro and C.~Perini,
  ``Coherent spin-networks,''
  Phys.\ Rev.\ D {\bf 82}, 024012 (2010)
  [arXiv:0912.4054 [gr-qc]].


\bibitem{Bojowald:2008zzb}
  M.~Bojowald,
  ``Loop Quantum Cosmology,''
  Living Rev.\ Rel.\  {\bf 11}, 4, (2008).
  \url{http://www.livingreviews.org/Articles/lrr-2008-4}.

\bibitem{Ashtekar:2011ni}
  A.~Ashtekar and P.~Singh,
  ``Loop Quantum Cosmology: A Status Report,''
  Class.\ Quant.\ Grav.\  {\bf 28}, 213001, (2011)
  [arXiv:1108.0893 [gr-qc]].

\bibitem{Bojowald:2011zzb}
  M.~Bojowald,
  ``Quantum Cosmology: A Fundamental Description of the Universe,''
  Lect.\ Notes Phys.\  {\bf 835}, 1, (2011).

\bibitem{Bojowald:2011book}
  M.~Bojowald,
  \emph{Canonical Gravity and Applications: Cosmology, Black Holes, and Quantum Gravity},
  (Cambridge University Press, Cambridge, 2011).


\bibitem{LaFave:1993zp}
  N.~J.~LaFave,
  ``A Step toward pregeometry. I: Ponzano-Regge spin networks and the origin of space-time structure in four-dimensions,''
  gr-qc/9310036.


\bibitem{Perez:2003vx}
  A.~Perez,
  ``Spin foam models for quantum gravity,''
  Class.\ Quant.\ Grav.\  {\bf 20}, R43 (2003)
  [gr-qc/0301113].

\bibitem{Oriti:2003wf}
  D.~Oriti,
  ``Spin foam models of quantum space-time,''
  gr-qc/0311066.

\bibitem{Perez:2004hj}
  A.~Perez,
  ``Introduction to loop quantum gravity and spin foams,''
  gr-qc/0409061.

\bibitem{Perez:2012wv}
  A.~Perez,
  ``The Spin Foam Approach to Quantum Gravity,''
  \emph{Living Rev.\ Rel.}\  {\bf 16}, 3, (2013).
  \url{http://www.livingreviews.org/Articles/lrr-2013-3}.
  [arXiv:1205.2019 [gr-qc]].


\bibitem{Rovelli:2010wq}
  C.~Rovelli,
  ``A new look at loop quantum gravity,''
  Class.\ Quant.\ Grav.\  {\bf 28}, 114005 (2011)
  [arXiv:1004.1780 [gr-qc]].

\bibitem{Rovelli:2011eq}
  C.~Rovelli,
  ``Zakopane lectures on loop gravity,''
  PoS QGQGS {\bf 2011}, 003 (2011)
  [arXiv:1102.3660 [gr-qc]].

\bibitem{Rovelli:2013ht}
  C.~Rovelli,
  ``Covariant loop gravity,''
  Lect.\ Notes Phys.\  {\bf 863}, 57 (2013).



\subsection*{Black hole thermodynamics}


\bibitem{Hawking:1971vc}
  S.~W.~Hawking,
  ``Black holes in general relativity,''
  Commun.\ Math.\ Phys.\  {\bf 25}, 152 (1972).

\bibitem{Bardeen:1973gs}
  J.~M.~Bardeen, B.~Carter and S.~W.~Hawking,
  ``The four laws of black hole mechanics,''
  Commun.\ Math.\ Phys.\  {\bf 31}, 161 (1973).


\bibitem{Wheeler's cup of tea}
  J.~A.~Wheeler and K.~Ford,
  ``Geons, Black Holes, and Quantum Foam,'' Chap.~4
  (W.~W.~Norton \& Company, New York, 1998).

\bibitem{Bekenstein:1972tm}
  J.~D.~Bekenstein,
  ``Black holes and the second law,''
  Lett.\ Nuovo Cim.\  {\bf 4}, 737 (1972).

\bibitem{Bekenstein:1973ur}
  J.~D.~Bekenstein,
  ``Black holes and entropy,''
  Phys.\ Rev.\ D {\bf 7}, 2333 (1973).

\bibitem{Bekenstein:1974ax}
  J.~D.~Bekenstein,
  ``Generalized second law of thermodynamics in black hole physics,''
  Phys.\ Rev.\ D {\bf 9}, 3292 (1974).




\bibitem{Hawking:1974rv}
  S.~W.~Hawking,
  Black hole explosions,
  \emph{Nature} {\bf 248}, 30 (1974).

\bibitem{Hawking:1974sw}
  S.~W.~Hawking,
  Particle Creation by Black Holes,
  \emph{Commun.\ Math.\ Phys.}\  {\bf 43} 199 (1975)
  [Erratum: \textit{ibid.}\  {\bf 46} 206 (1976)].


\bibitem{Unruh:1976db}
  W.~G.~Unruh,
  ``Notes on black hole evaporation,''
  Phys.\ Rev.\ D {\bf 14}, 870 (1976).

\bibitem{Bisognano:1976za}
  J.~J.~Bisognano and E.~H.~Wichmann,
  ``On the Duality Condition for Quantum Fields,''
  J.\ Math.\ Phys.\  {\bf 17}, 303 (1976).

\bibitem{DeWitt:Quantum Gravity}
  B.~S.~DeWitt,
  ``Quantum gravity: The new synthesis,''
  in \emph{General Relativity: An Einstein Centenary Survey},
  eds.\ S.~W.~Hawking and W.~Israel
  (Cambridge University Press, Cambridge, 1979).


\bibitem{Gibbons:1976es}
  G.~W.~Gibbons and M.~J.~Perry,
  ``Black Holes in Thermal Equilibrium,''
  Phys.\ Rev.\ Lett.\  {\bf 36}, 985 (1976).

\bibitem{Gibbons:1978}
  G.~W.~Gibbons and M.~J.~Perry,
  ``Black hole and thermal Green functions,''
  Proc.\ Roy.\ Soc.\ Lond.\ A {\bf 358} 467–494 (1978).

\bibitem{Gibbons:1976ue}
  G.~W.~Gibbons and S.~W.~Hawking,
  ``Action Integrals and Partition Functions in Quantum Gravity,''
  Phys.\ Rev.\ D {\bf 15}, 2752 (1977).


\bibitem{Carlip:1993sa}
  S.~Carlip and C.~Teitelboim,
  ``The off-shell black hole,''
  Class.\ Quant.\ Grav.\  {\bf 12} (1995) 1699
  [gr-qc/9312002].


\bibitem{Parikh:1999mf}
  M.~K.~Parikh and F.~Wilczek,
  ``Hawking radiation as tunneling,''
  Phys.\ Rev.\ Lett.\  {\bf 85}, 5042 (2000)
  [hep-th/9907001].

\bibitem{Vanzo:2011wq}
  L.~Vanzo, G.~Acquaviva and R.~Di Criscienzo,
  ``Tunnelling Methods and Hawking's radiation: achievements and prospects,''
  Class.\ Quant.\ Grav.\  {\bf 28}, 183001 (2011)
  [arXiv:1106.4153 [gr-qc]].


\bibitem{Robinson:2005pd}
  S.~P.~Robinson and F.~Wilczek,
  ``A Relationship between Hawking radiation and gravitational anomalies,''
  Phys.\ Rev.\ Lett.\  {\bf 95}, 011303 (2005)
  [gr-qc/0502074].

\bibitem{Iso:2006wa}
  S.~Iso, H.~Umetsu and F.~Wilczek,
  ``Hawking radiation from charged black holes via gauge and gravitational anomalies,''
  Phys.\ Rev.\ Lett.\  {\bf 96}, 151302 (2006)
  [hep-th/0602146].

\bibitem{Iso:2006ut}
  S.~Iso, H.~Umetsu and F.~Wilczek,
  ``Anomalies, Hawking radiations and regularity in rotating black holes,''
  Phys.\ Rev.\ D {\bf 74}, 044017 (2006)
  [hep-th/0606018].

\bibitem{Iso:2007nf}
  S.~Iso, T.~Morita and H.~Umetsu,
  ``Hawking radiation via higher-spin gauge anomalies,''
  Phys.\ Rev.\ D {\bf 77}, 045007 (2008)
  [arXiv:0710.0456 [hep-th]].

\bibitem{Bonora:2008nk}
  L.~Bonora, M.~Cvitan, S.~Pallua and I.~Smoli\'{c},
  ``Hawking Fluxes, W(infinity) Algebra and Anomalies,''
  JHEP {\bf 0812}, 021 (2008)
  [arXiv:0808.2360 [hep-th]].

\bibitem{Bonora:2009tw}
  L.~Bonora, M.~Cvitan, S.~Pallua and I.~Smoli\'{c},
  ``Hawking fluxes, Fermionic currents, W(1+infinity) algebra and anomalies,''
  Phys.\ Rev.\ D {\bf 80}, 084034 (2009)
  [arXiv:0907.3722 [hep-th]].


\bibitem{Garfinkle:1993xk}
  D.~Garfinkle, S.~B.~Giddings and A.~Strominger,
  ``Entropy in black hole pair production,''
  Phys.\ Rev.\ D {\bf 49}, 958 (1994)
  [gr-qc/9306023].

\bibitem{Brown:1994um}
  J.~D.~Brown,
  ``Black hole pair creation and the entropy factor,''
  Phys.\ Rev.\ D {\bf 51}, 5725 (1995)
  [gr-qc/9412018].

\bibitem{Dowker:1993bt}
  F.~Dowker, J.~P.~Gauntlett, D.~A.~Kastor and J.~H.~Traschen,
  ``Pair creation of dilaton black holes,''
  Phys.\ Rev.\ D {\bf 49}, 2909 (1994)
  [hep-th/9309075].

\bibitem{Mann:1995vb}
  R.~B.~Mann and S.~F.~Ross,
  ``Cosmological production of charged black hole pairs,''
  Phys.\ Rev.\ D {\bf 52}, 2254 (1995)
  [gr-qc/9504015].


\bibitem{Strominger:1996sh}
  A.~Strominger and C.~Vafa,
  ``Microscopic origin of the Bekenstein-Hawking entropy,''
  Phys.\ Lett.\ B {\bf 379}, 99 (1996)
  [hep-th/9601029].

\bibitem{Peet:2000hn}
  A.~W.~Peet,
  ``TASI lectures on black holes in string theory,''
  in \emph{TASI 99: Strings, Branes and Gravity},
  eds.\ J.~Harvey, S.~Kachru and E.~Silverstein
  (World Scientiﬁc, Singapore, 2001)
  [hep-th/0008241].

\bibitem{Das:2000su}
  S.~R.~Das and S.~D.~Mathur,
  ``The quantum physics of black holes: Results from string theory,''
  Ann.\ Rev.\ Nucl.\ Part.\ Sci.\  {\bf 50}, 153 (2000)
  [gr-qc/0105063].

\bibitem{Mathur:2005zp}
  S.~D.~Mathur,
  ``The Fuzzball proposal for black holes: An Elementary review,''
  Fortsch.\ Phys.\  {\bf 53}, 793 (2005)
  [hep-th/0502050].

\bibitem{Mathur:2005ai}
  S.~D.~Mathur,
  ``The Quantum structure of black holes,''
  Class.\ Quant.\ Grav.\  {\bf 23}, R115 (2006)
  [hep-th/0510180].

\bibitem{Banados:1992wn}
  M.~Banados, C.~Teitelboim and J.~Zanelli,
  ``The Black hole in three-dimensional space-time,''
  Phys.\ Rev.\ Lett.\  {\bf 69}, 1849 (1992)
  [hep-th/9204099].

\bibitem{Strominger:1997eq}
  A.~Strominger,
  ``Black hole entropy from near horizon microstates,''
  JHEP {\bf 9802}, 009 (1998)
  [hep-th/9712251].

\bibitem{Birmingham:1998jt}
  D.~Birmingham, I.~Sachs and S.~Sen,
  ``Entropy of three-dimensional black holes in string theory,''
  Phys.\ Lett.\ B {\bf 424}, 275 (1998)
  [hep-th/9801019].

\bibitem{Skenderis:1999bs}
  K.~Skenderis,
  ``Black holes and branes in string theory,''
  Lect.\ Notes Phys.\  {\bf 541}, 325 (2000)
  [hep-th/9901050].



\bibitem{Krasnov:1996tb}
  K.~V.~Krasnov,
  ``Counting surface states in the loop quantum gravity,''
  Phys.\ Rev.\ D {\bf 55}, 3505 (1997)
  [gr-qc/9603025].

\bibitem{Ashtekar:1997yu}
  A.~Ashtekar, J.~Baez, A.~Corichi and K.~Krasnov,
  ``Quantum geometry and black hole entropy,''
  Phys.\ Rev.\ Lett.\  {\bf 80}, 904 (1998)
  [gr-qc/9710007].

\bibitem{Ashtekar:1999wa}
  A.~Ashtekar, A.~Corichi and K.~Krasnov,
  ``Isolated horizons: The Classical phase space,''
  Adv.\ Theor.\ Math.\ Phys.\  {\bf 3}, 419 (1999)
  [gr-qc/9905089].

\bibitem{Ashtekar:2000eq}
  A.~Ashtekar, J.~C.~Baez and K.~Krasnov,
  ``Quantum geometry of isolated horizons and black hole entropy,''
  Adv.\ Theor.\ Math.\ Phys.\  {\bf 4}, 1 (2000)
  [gr-qc/0005126].

\bibitem{Ashtekar:2000hw}
  A.~Ashtekar, S.~Fairhurst and B.~Krishnan,
  ``Isolated horizons: Hamiltonian evolution and the first law,''
  Phys.\ Rev.\ D {\bf 62}, 104025 (2000)
  [gr-qc/0005083].

\bibitem{Ashtekar:2001is}
  A.~Ashtekar, C.~Beetle and J.~Lewandowski,
  ``Mechanics of rotating isolated horizons,''
  Phys.\ Rev.\ D {\bf 64}, 044016 (2001)
  [gr-qc/0103026].

\bibitem{Ashtekar:2001jb}
  A.~Ashtekar, C.~Beetle and J.~Lewandowski,
  ``Geometry of generic isolated horizons,''
  Class.\ Quant.\ Grav.\  {\bf 19}, 1195 (2002)
  [gr-qc/0111067].

\bibitem{Ashtekar:2003jh}
  A.~Ashtekar, A.~Corichi and D.~Sudarsky,
  ``Nonminimally coupled scalar fields and isolated horizons,''
  Class.\ Quant.\ Grav.\  {\bf 20}, 3413 (2003)
  [gr-qc/0305044].

\bibitem{Ashtekar:2003zx}
  A.~Ashtekar and A.~Corichi,
  ``Nonminimal couplings, quantum geometry and black hole entropy,''
  Class.\ Quant.\ Grav.\  {\bf 20}, 4473 (2003)
  [gr-qc/0305082].

\bibitem{Ashtekar:2004nd}
  A.~Ashtekar, J.~Engle and C.~Van Den Broeck,
  ``Quantum horizons and black hole entropy: Inclusion of distortion and rotation,''
  Class.\ Quant.\ Grav.\  {\bf 22}, L27 (2005)
  [gr-qc/0412003].

\bibitem{Domagala:2004jt}
  M.~Domaga{\l}a and J.~Lewandowski,
  ``Black hole entropy from quantum geometry,''
  Class.\ Quant.\ Grav.\  {\bf 21}, 5233 (2004)
  [gr-qc/0407051].

\bibitem{Meissner:2004ju}
  K.~A.~Meissner,
  ``Black hole entropy in loop quantum gravity,''
  Class.\ Quant.\ Grav.\  {\bf 21}, 5245 (2004)
  [gr-qc/0407052].

\bibitem{Corichi:2006bs}
  A.~Corichi, J.~Diaz-Polo and E.~Fernandez-Borja,
  ``Quantum geometry and microscopic black hole entropy,''
  Class.\ Quant.\ Grav.\  {\bf 24}, 243 (2007)
  [gr-qc/0605014].

\bibitem{Corichi:2006wn}
  A.~Corichi, J.~Diaz-Polo and E.~Fernandez-Borja,
  ``Black hole entropy quantization,''
  Phys.\ Rev.\ Lett.\  {\bf 98}, 181301 (2007)
  [gr-qc/0609122].

\bibitem{Agullo:2010zz}
  I.~Agullo, J.~Fernando Barbero, E.~F.~Borja, J.~Diaz-Polo and E.~J.~S.~Villasenor,
  ``Detailed black hole state counting in loop quantum gravity,''
  Phys.\ Rev.\ D {\bf 82}, 084029 (2010)
  [arXiv:1101.3660 [gr-qc]].

\bibitem{BarberoG.:2012ae}
  J.~F.~Barbero G., J.~Lewandowski and E.~J.~S.~Villasenor,
  ``Quantum isolated horizons and black hole entropy,''
  PoS QGQGS {\bf 2011}, 023 (2011)
  [arXiv:1203.0174 [gr-qc]].

\bibitem{Bekenstein:1995ju}
  J.~D.~Bekenstein and V.~F.~Mukhanov,
  ``Spectroscopy of the quantum black hole,''
  Phys.\ Lett.\ B {\bf 360}, 7 (1995)
  [gr-qc/9505012].

\bibitem{Engle:2009vc}
  J.~Engle, A.~Perez and K.~Noui,
  ``Black hole entropy and $SU(2)$ Chern-Simons theory,''
  Phys.\ Rev.\ Lett.\  {\bf 105}, 031302 (2010)
  [arXiv:0905.3168 [gr-qc]].

\bibitem{Engle:2010kt}
  J.~Engle, K.~Noui, A.~Perez and D.~Pranzetti,
  ``Black hole entropy from an $SU(2)$-invariant formulation of Type I isolated horizons,''
  Phys.\ Rev.\ D {\bf 82}, 044050 (2010)
  [arXiv:1006.0634 [gr-qc]].

\bibitem{Engle:2011vf}
  J.~Engle, K.~Noui, A.~Perez and D.~Pranzetti,
  ``The $SU(2)$ Black Hole entropy revisited,''
  JHEP {\bf 1105}, 016 (2011)
  [arXiv:1103.2723 [gr-qc]].


\bibitem{Sorkin:2014kta}
  R.~D.~Sorkin,
  ``On the Entropy of the Vacuum outside a Horizon,''
  Tenth International Conference on General Relativity and Gravitation
  (held Padova, 4-9 July, 1983), Contributed Papers, vol.~II, pp.\
  734-736
  [arXiv:1402.3589 [gr-qc]].

\bibitem{Bombelli:1986rw}
  L.~Bombelli, R.~K.~Koul, J.~Lee and R.~D.~Sorkin,
  ``A Quantum Source of Entropy for Black Holes,''
  Phys.\ Rev.\ D {\bf 34}, 373 (1986).

\bibitem{Srednicki:1993im}
  M.~Srednicki,
  ``Entropy and area,''
  Phys.\ Rev.\ Lett.\  {\bf 71}, 666 (1993)
  [hep-th/9303048].

\bibitem{Ryu:2006bv}
  S.~Ryu and T.~Takayanagi,
  ``Holographic derivation of entanglement entropy from AdS/CFT,''
  Phys.\ Rev.\ Lett.\  {\bf 96}, 181602 (2006)
  [hep-th/0603001].

\bibitem{Hubeny:2007xt}
  V.~E.~Hubeny, M.~Rangamani and T.~Takayanagi,
  ``A Covariant holographic entanglement entropy proposal,''
  JHEP {\bf 0707}, 062 (2007)
  [arXiv:0705.0016 [hep-th]].


\bibitem{Sakharov:1967pk}
  A.~D.~Sakharov,
  ``Vacuum quantum fluctuations in curved space and the theory of gravitation,''
  Sov.\ Phys.\ Dokl.\  {\bf 12}, 1040 (1968)
  [Dokl.\ Akad.\ Nauk Ser.\ Fiz.\  {\bf 177}, 70 (1967)]
  [Sov.\ Phys.\ Usp.\  {\bf 34}, 394 (1991)]
  [Gen.\ Rel.\ Grav.\  {\bf 32}, 365 (2000)].

\bibitem{Frolov:1996aj}
  V.~P.~Frolov, D.~V.~Fursaev and A.~I.~Zelnikov,
  ``Statistical origin of black hole entropy in induced gravity,''
  Nucl.\ Phys.\ B {\bf 486}, 339 (1997)
  [hep-th/9607104].

\bibitem{Frolov:1997up}
  V.~P.~Frolov and D.~V.~Fursaev,
  Phys.\ Rev.\ D {\bf 56}, 2212 (1997)
  [hep-th/9703178].


\bibitem{Carlip:2000nv}
  S.~Carlip,
  ``Logarithmic corrections to black hole entropy from the Cardy formula,''
  Class.\ Quant.\ Grav.\  {\bf 17}, 4175 (2000)
  [gr-qc/0005017].


\bibitem{Mathur:2008wi}
  S.~D.~Mathur,
  ``What Exactly is the Information Paradox?,''
  Lect.\ Notes Phys.\  {\bf 769}, 3 (2009)
  [arXiv:0803.2030 [hep-th]].


\bibitem{Carlip:2014pma}
  S.~Carlip,
  ``Black Hole Thermodynamics,''
  Int.\ J.\ Mod.\ Phys.\ D {\bf 23}, 1430023 (2014)
  [arXiv:1410.1486 [gr-qc]].


\subsection*{Quantum gravity phenomenology}

\bibitem{Kostelecky:1988zi}
  V.~A.~Kostelecky and S.~Samuel,
  ``Spontaneous Breaking of Lorentz Symmetry in String Theory,''
  Phys.\ Rev.\ D {\bf 39}, 683 (1989).

\bibitem{Burgess:2002tb}
  C.~P.~Burgess, J.~M.~Cline, E.~Filotas, J.~Matias and G.~D.~Moore,
  ``Loop generated bounds on changes to the graviton dispersion relation,''
  JHEP {\bf 0203}, 043 (2002)
  [hep-ph/0201082].

\bibitem{Gambini:1998it}
  R.~Gambini and J.~Pullin,
  ``Nonstandard optics from quantum space-time,''
  Phys.\ Rev.\ D {\bf 59}, 124021 (1999)
  [gr-qc/9809038].

\bibitem{Rovelli:2010ed}
  C.~Rovelli and S.~Speziale,
  ``Lorentz covariance of loop quantum gravity,''
  Phys.\ Rev.\ D {\bf 83}, 104029 (2011)
  [arXiv:1012.1739 [gr-qc]].

\bibitem{Lukierski:1993wx}
  J.~Lukierski, H.~Ruegg and W.~J.~Zakrzewski,
  ``Classical quantum mechanics of free kappa relativistic systems,''
  Annals Phys.\  {\bf 243}, 90 (1995)
  [hep-th/9312153].

\bibitem{AmelinoCamelia:1999pm}
  G.~Amelino-Camelia and S.~Majid,
  ``Waves on noncommutative space-time and gamma-ray bursts,''
  Int.\ J.\ Mod.\ Phys.\ A {\bf 15} (2000) 4301
  [hep-th/9907110].

\bibitem{Carroll:2001ws}
  S.~M.~Carroll, J.~A.~Harvey, V.~A.~Kostelecky, C.~D.~Lane and T.~Okamoto,
  ``Noncommutative field theory and Lorentz violation,''
  Phys.\ Rev.\ Lett.\  {\bf 87}, 141601 (2001)
  [hep-th/0105082].

\bibitem{Barcelo:2005fc}
  C.~Barcelo, S.~Liberati and M.~Visser,
  ``Analogue gravity,''
  Living Rev.\ Rel.\  {\bf 8}, 12 (2005)
  [Living Rev.\ Rel.\  {\bf 14}, 3 (2011)]
  \url{http://relativity.livingreviews.org/Articles/lrr-2011-3/}
  [gr-qc/0505065].

\bibitem{Ni:2014qfa}
  W.~T.~Ni,
  ``Spacetime structure and asymmetric metric from the premetric formulation of electromagnetism,''
  Phys.\ Lett.\ A {\bf 379}, 1297 (2015)
  [arXiv:1411.0460 [gr-qc]].

\bibitem{Ni:2014cca}
  W.~T.~Ni,
  ``Dilaton field and cosmic wave propagation,''
  Phys.\ Lett.\ A {\bf 378}, 3413 (2014)
  [arXiv:1410.0126 [gr-qc]].

\bibitem{Alighieri:2015hta}
  S.~d.~S.~Alighieri,
  ``Cosmic Polarization Rotation: an Astrophysical Test of Fundamental Physics,''
  Int.\ J.\ Mod.\ Phys.\ D {\bf 24}, 0016 (2015)
  [arXiv:1501.06460 [astro-ph.CO]].

\bibitem{Mei:2014iaa}
  H.~H.~Mei, W.~T.~Ni, W.~P.~Pan, L.~Xu and S.~d.~S.~Alighieri,
  ``New constraints on cosmic polarization rotation from the ACTPol cosmic microwave background B-Mode polarization observation and the BICEP2 constraint update,''
  Astrophys.\ J.\  {\bf 805}, no. 2, 107 (2015)
  [arXiv:1412.8569 [astro-ph.CO]].

\bibitem{Kotera:2011cp}
  K.~Kotera and A.~V.~Olinto,
  ``The Astrophysics of Ultrahigh Energy Cosmic Rays,''
  Ann.\ Rev.\ Astron.\ Astrophys.\  {\bf 49}, 119 (2011)
  [arXiv:1101.4256 [astro-ph.HE]].

\bibitem{Yamamoto:2007xj}
  T.~Yamamoto [for the Pierre Auger Collaboration],
  ``The UHECR spectrum measured at the Pierre Auger Observatory and its astrophysical implications,''
  arXiv:0707.2638 [astro-ph].

\bibitem{Stecker:2009hj}
  F.~W.~Stecker and S.~T.~Scully,
  ``Searching for New Physics with Ultrahigh Energy Cosmic Rays,''
  New J.\ Phys.\  {\bf 11}, 085003 (2009)
  [arXiv:0906.1735 [astro-ph.HE]].

\bibitem{Liberati:2011bp}
  S.~Liberati and L.~Maccione,
  ``Quantum Gravity phenomenology: achievements and challenges,''
  J.\ Phys.\ Conf.\ Ser.\  {\bf 314}, 012007 (2011)
  [arXiv:1105.6234 [astro-ph.HE]].

\bibitem{AmelinoCamelia:2010zf}
  G.~Amelino-Camelia, A.~Marciano, M.~Matassa and G.~Rosati,
  ``Testing quantum-spacetime relativity with gamma-ray telescopes,''
  arXiv:1006.0007 [astro-ph.HE].

\bibitem{AmelinoCamelia:2009zzb}
  G.~Amelino-Camelia, C.~Laemmerzahl, F.~Mercati and G.~M.~Tino,
  ``Constraining the Energy-Momentum Dispersion Relation with Planck-Scale Sensitivity Using Cold Atoms,''
  Phys.\ Rev.\ Lett.\  {\bf 103}, 171302 (2009)
  [arXiv:0911.1020 [gr-qc]].

\bibitem{Mercati:2010au}
  F.~Mercati, D.~Mazon, G.~Amelino-Camelia, J.~M.~Carmona, J.~L.~Cortes, J.~Indurain, C.~Laemmerzahl and G.~M.~Tino,
  ``Probing the quantum-gravity realm with slow atoms,''
  Class.\ Quant.\ Grav.\  {\bf 27}, 215003 (2010)
  [arXiv:1004.0847 [gr-qc]].

\bibitem{Damour:1992kf}
  T.~Damour and K.~Nordtvedt,
  ``General relativity as a cosmological attractor of tensor scalar theories,''
  Phys.\ Rev.\ Lett.\  {\bf 70}, 2217 (1993).

\bibitem{Dvali:2000hr}
  G.~R.~Dvali, G.~Gabadadze and M.~Porrati,
  ``4-D gravity on a brane in 5-D Minkowski space,''
  Phys.\ Lett.\ B {\bf 485}, 208 (2000)
  [hep-th/0005016].

\bibitem{deRham:2014zqa}
  C.~de Rham,
  ``Massive Gravity,''
  Living Rev.\ Rel.\  {\bf 17}, 7 (2014)
  \url{http://relativity.livingreviews.org/Articles/lrr-2014-7/}
  [arXiv:1401.4173 [hep-th]].

\bibitem{Ni:2008bj}
  W.~T.~Ni,
  ``Super-ASTROD: Probing primordial gravitational waves and mapping the outer solar system,''
  Class.\ Quant.\ Grav.\  {\bf 26}, 075021 (2009)
  [arXiv:0812.0887 [astro-ph]].

\bibitem{Braxmaier:2011ai}
  C.~Braxmaier {\it et al.} [ASTROD Collaboration],
  ``Astrodynamical Space Test of Relativity using Optical Devices I (ASTROD I) - A class-M fundamental physics mission proposal for Cosmic Vision 2015-2025: 2010 Update,''
  Exper.\ Astron.\  {\bf 34}, 181 (2012)
  [arXiv:1104.0060 [gr-qc]].


\bibitem{Gair:2012nm}
  J.~R.~Gair, M.~Vallisneri, S.~L.~Larson and J.~G.~Baker,
  ``Testing General Relativity with Low-Frequency, Space-Based Gravitational-Wave Detectors,''
  Living Rev.\ Rel.\  {\bf 16}, 7 (2013)
  \url{http://relativity.livingreviews.org/Articles/lrr-2013-7/}
  [arXiv:1212.5575 [gr-qc]].

\bibitem{Lammerzahl:2004tc}
  C.~L\"{a}mmerzahl,
  ``General relativity in space and sensitive tests of the equivalence principle,''
  gr-qc/0402122.

\bibitem{Ashtekar:2013xka}
  A.~Ashtekar,
  ``Loop Quantum Gravity and the The Planck Regime of Cosmology,''
  arXiv:1303.4989 [gr-qc].

\bibitem{Grain:2009kw}
  J.~Grain and A.~Barrau,
  ``Cosmological footprints of loop quantum gravity,''
  Phys.\ Rev.\ Lett.\  {\bf 102}, 081301 (2009)
  [arXiv:0902.0145 [gr-qc]].

\bibitem{Barrau:2009fz}
  A.~Barrau,
  ``Loop quantum gravity and the CMB: Toward pre-Big Bounce cosmology,''
  arXiv:0911.3745 [gr-qc].

\bibitem{Mielczarek:2010bh}
  J.~Mielczarek, T.~Cailleteau, J.~Grain and A.~Barrau,
  ``Inflation in loop quantum cosmology: dynamics and spectrum of gravitational waves,''
  Phys.\ Rev.\ D {\bf 81}, 104049 (2010)
  [arXiv:1003.4660 [gr-qc]].

\bibitem{Grain:2010yv}
  J.~Grain, A.~Barrau, T.~Cailleteau and J.~Mielczarek,
 `` Observing the Big Bounce with Tensor Modes in the Cosmic Microwave Background: Phenomenology and Fundamental LQC Parameters,''
  Phys.\ Rev.\ D {\bf 82}, 123520 (2010)
  [arXiv:1011.1811 [astro-ph.CO]].

\bibitem{Barrau:2010nd}
  A.~Barrau,
  ``Inflation and Loop Quantum Cosmology,''
  PoS ICHEP {\bf 2010}, 461 (2010)
  [arXiv:1011.5516 [gr-qc]].

\bibitem{Agullo:2013ai}
  I.~Agullo, A.~Ashtekar and W.~Nelson,
  ``The pre-inflationary dynamics of loop quantum cosmology: Confronting quantum gravity with observations,''
  Class.\ Quant.\ Grav.\  {\bf 30}, 085014 (2013)
  [arXiv:1302.0254 [gr-qc]].

\bibitem{Ade:2015tva}
  P.~A.~R.~Ade {\it et al.}  [BICEP2 and Planck Collaborations],
  ``Joint Analysis of BICEP2/$Keck  Array$ and $Planck$ Data,''
  Phys.\ Rev.\ Lett.\  {\bf 114}, no. 10, 101301 (2015)
  [arXiv:1502.00612 [astro-ph.CO]].

\bibitem{AmelinoCamelia:2008qg}
  G.~Amelino-Camelia,
  ``Quantum-Spacetime Phenomenology,''
  Living Rev.\ Rel.\  {\bf 16}, 5 (2013)
  \url{http://relativity.livingreviews.org/Articles/lrr-2013-5/}
  [arXiv:0806.0339 [gr-qc]].

\bibitem{Smolin:2005re}
  L.~Smolin,
  ``Loop quantum gravity and Planck scale phenomenology,''
  Lect.\ Notes Phys.\  {\bf 669}, 363 (2005).


\end{thebibliography}
\end{document}